\documentclass[11pt,english,epsf]{article}
\usepackage{amsmath}
\usepackage{amsfonts}
\usepackage{amssymb}
\usepackage{graphicx}
\usepackage{color}
\usepackage{comment}

\newtheorem{theorem}{Theorem}
\newtheorem{lemma}{Lemma}
\newtheorem{corollary}{Corollary}
\DeclareMathOperator*{\E}{\mathbb{E}}
\newcommand {\lexe} {\stackrel{\cdot} {\le}}

\newcommand {\dfn} {\stackrel{\Delta} {=}}
\newcommand {\exe} {\stackrel{\cdot} {=}}

\newcommand {\tR} {\tilde{R}}

\newcommand {\bu} {\mbox{\boldmath $u$}}
\newcommand {\bv} {\mbox{\boldmath $v$}}

\newcommand {\bx} {\mbox{\boldmath $x$}}
\newcommand {\by} {\mbox{\boldmath $y$}}

\newcommand {\bE} {\mbox{\boldmath $E$}}

\newcommand {\bV} {\mbox{\boldmath $V$}}

\newcommand {\bX} {\mbox{\boldmath $X$}}
\newcommand {\bY} {\mbox{\boldmath $Y$}}

\newcommand{\calA}{{\cal A}}
\newcommand{\calB}{{\cal B}}

\newcommand{\calE}{{\cal E}}

\newcommand{\calI}{{\cal I}}

\newcommand{\calS}{{\cal S}}
\newcommand{\calT}{{\cal T}}
\newcommand{\calU}{{\cal U}}
\newcommand{\calV}{{\cal V}}

\newcommand{\calX}{{\cal X}}
\newcommand{\calY}{{\cal Y}}

\newcommand {\tbx} {\tilde{\mbox{\boldmath $x$}}}
\newcommand {\hbx} {\hat{\mbox{\boldmath $x$}}}

\newcommand {\tby} {\tilde{\mbox{\boldmath $y$}}}
\newcommand {\hby} {\hat{\mbox{\boldmath $y$}}}

\newcommand {\hH} {\hat{H}}
\newcommand {\hP} {\hat{P}}
\newcommand {\hI} {\hat{I}}
\allowdisplaybreaks

\begin{document}
\thispagestyle{empty}
\title{On More General Distributions of
Random Binning\\ 
for Slepian-Wolf Encoding\thanks{
	This research is partially supported by the Israel Science Foundation (ISF), grant no.\ 137/18.}}
\author{Neri Merhav}
\date{}
\maketitle

\begin{center}
The Andrew \& Erna Viterbi Faculty of Electrical Engineering\\
Technion - Israel Institute of Technology \\
Technion City, Haifa 32000, ISRAEL \\
E--mail: {\tt merhav@ee.technion.ac.il}\\
\end{center}
\vspace{1.5\baselineskip}
\setlength{\baselineskip}{1.5\baselineskip}

\begin{abstract}
We consider the problem of (almost) lossless separate encodings
and joint decoding of two correlated discrete memoryless sources (DMSs), 
that is, the well--known Slepian-Wolf (S-W) source coding problem. 
Traditionally, ensembles of S--W codes are defined
such that every bin of each $n$--vector of each source is 
randomly drawn under the uniform distribution across the
sets $\{0,1,\ldots,2^{nR_X}-1\}$
and $\{0,1,\ldots,2^{nR_Y}-1\}$, 
where $R_X$ and $R_Y$ are the coding rates of the two sources, $X$ and $Y$, respectively. 
In a few more recent works, where only one source, say, $X$, 
is compressed and the other one, $Y$, serves as
side information available at the decoder,
the scope is extended to variable--rate S--W (VRSW) codes,
where the rate is allowed to depend on the type class of 
the source string, but still, the random--binning distribution is assumed uniform
within the corresponding, type--dependent, bin index set. 

In this expository work, we investigate the 
role of the uniformity of the random binning 
distribution from the perspective of the trade-off between 
the reliability (defined in terms of the error exponent) 
and the compression performance (measured from the viewpoint of the source coding exponent).
To this end, we study a much wider class of random--binning
distributions, which includes the ensemble of VRSW codes 
as a special case, but it also goes considerably beyond. 
We first show that, with the exception of some pathological cases, 
the smaller ensemble, of VRSW codes, is as good as the larger ensemble in terms 
the trade--off between the error exponent and the source coding exponent. 
Notwithstanding this finding, the wider class of ensembles proposed
is motivated in two ways. The first is that it outperforms 
VRSW codes in the above--mentioned pathological cases, 
and the second is that it allows robustness:
in the event of a system failure that causes unavailability 
of the compressed bit--stream from one of the sources, 
it still allows reconstruction of the other source within some controllable distortion.\\

\noindent
{\bf Index Terms:} Slepian--Wolf coding, random binning, universal decoding, error exponent, 
source coding exponent, variable--rate coding.
\end{abstract}

\newpage
\section{Introduction}

The problem of (almost) loss separate source coding and joint decoding of two correlated
sources, i.e., the well
known Slepian--Wolf (S-W) coding problem, has received a vast level of attention ever since
the celebrated article by Slepian and Wolf \cite{SW73} was published, nearly five decades ago. 

Ensembles of S--W codes have almost exclusively
been defined by independent, random selection of a bin indices 
under the {\it uniform distribution}, 
for each and every possible source
vector. In particular, denoting generically the two sources by $X$ and $Y$, each $n$--vector
of $X$ (resp.\ $Y$) is mapped into a randomly selected bin in $\{0,1,\ldots,2^{nR_X}\}$ (resp.\
$\{0,1,\ldots,2^{nR_Y}\}$), which is described using
$nR_X$ (resp.\ $nR_Y$) bits. To the best knowledge of the author, the 
only exception to this rule
is in the notion of variable--rate S--W (VRSW) 
codes, see, e.g., \cite{CHJ09}, \cite{CHJL08a}, 
\cite{CHJL08b}, \cite{biomet}, \cite{WM15}. In those works, the simpler version
of S--W setting was considered, where only one source, say, $X$, is compressed, whereas the
other one, $Y$, serves as side information available at the decoder, and
the coding rate for $X$, is allowed to vary with dependence 
on the type class of the source sequence, 
$\bx=(x_1,\ldots,x_n)$, or equivalently, 
the empirical distribution of the source sequence, $\hP_{\bx}$. As was shown in \cite{WM15}, 
VRSW codes offer considerable improvement relative to their fixed--rate counterparts, 
in terms of achievable trade--offs between the error exponent 
and the source coding exponent. Moreover, the optimal rate function, $R(\hP_{\bx})$,
was characterized in \cite{WM15}.
However, here too, within each and every type class of source
sequences, the ensemble of codes is defined by independent 
random selection bin assignment under the uniform distribution over
$\{0,1,\ldots,2^{nR(\hP_{\bx})}-1\}$.

It is quite natural to ask, at this point, 
why is the uniform distribution being used so exclusively in the context of S--W
random binning, whereas in other contexts of ensembles of codes 
(for both source coding and channel coding),
general random coding distributions are used, 
which depend on the underlying source, or channel. 
A simple, intuitive answer could be that the bins are 
randomly drawn in the ``compressed domain'':
after optimal compression, strings of bits form a 
binary symmetric source (or at least, nearly so), and therefore, they are distributed uniformly.
On the face of it, such an answer sounds appealing and satisfactory.
Indeed, this uniform distribution is good enough for proving 
the S--W coding theorem, which is compatible with the converse theorem.
However, as will be demonstrated shortly, this kind of explanation 
is somewhat too simplistic when it comes to more refined figures of merit, like
the error exponent and the source coding exponent. 
As will be demonstrated shortly, the optimality of the uniform random--binning
distribution is not quite obvious when one examines 
the problem through the lens of those figures of merit.

In view of this background, our aim, in this expository paper, is 
to study this question with some more detail. We carry out this study by
investigating a considerably wider class of bin--assignment 
distributions from the viewpoint of
achievable trade-offs between the error exponent and the source coding exponent.
In particular, we consider an ensemble model where 
source $n$--vectors, $\bx$ and $\by$, are mapped 
into `bins' that are represented by other finite--alphabet $n$--vectors,
$\bu=(u_1,\ldots,u_n)\in\calU^n$ and $\bv=(v_1,\ldots,v_n)\in\calV^n$, respectively, that are drawn under 
conditional probability distributions, $A(\bu|\bx)$ and $B(\bv|\by)$, which have a simple
structure: $A(\bu|\bx)$ (resp.\ $B(\bv|\by)$) depends on $\bx$ and $\bu$ (resp.\
$\by$ and $\bv$) only via their empirical joint distribution, $\hP_{\bu\bx}$ (resp.\
$\hP_{\bv\by}$). More precisely,
\begin{equation}
\label{genericW}
A(\bu|\bx)\exe\exp\{-nF(\hP_{\bu\bx})\},~~~~~
B(\bv|\by)\exe\exp\{-nG(\hP_{\bv\by})\},
\end{equation}
where $\exe$ stands for asymptotic equality in the 
exponential order (to be formally defined in the sequel) and $F$ and $G$ 
are non-negative functions
of the respective empirical joint distributions.
The outputs, $\bu$ and $\bv$, of these mappings, are then losslessly compressed 
and sent to the
decoder, which in turn, first reconstructs $\bu$ and $\bv$, and then 
selects the most likely pair of source vectors among all 
those mapped to the given $\bu$ and $\bv$.

The motivation for studying the class of conditional distributions (\ref{genericW}), henceforth referred to as {\it random--binning channels} (RBCs), 
is that, on the one hand, thanks to their structure, they are 
still amenable to analysis using the standard method
of types \cite{CK81}, and on the other hand, they are general enough to 
cover a large variety of relevant special cases, 
such as memoryless sources and their mixtures,
memoryless channels and their mixtures, 
uniform distributions within single (or several) type classes, 
or conditional type classes, etc.
Obviously, the ordinary ensemble of S--W codes corresponds 
to the special case where $F(\hP_{\bu\bx})\equiv R_X$, $G(\hP_{\bv\by})\equiv R_Y$,
and the alphabets, $\calU$ and $\calV$, of each $u_i$ and $v_i$, are of 
sizes $2^{R_X}$ and $2^{R_Y}$, respectively.\footnote{If $2^{R_X}$ is not integer, 
the rate $R_X$ can be approximated arbitrarily closely by alphabet
extensions. A similar comment applies, of course, also to $R_Y$.} 
VRSW codes can be viewed in this formal framework too: for given binning alphabets, $\calU$ and $\calV$, one may define
$F(\hP_{\bu\bx})=H(\hP_{\bu})$ and
$G(\hP_{\bv\by})=H(\hP_{\bv})$, i.e.,
the entropies associated with certain empirical distributions, $\hP_{\bu}$ and $\hP_{\bv}$, 
that may depend on $\hP_{\bx}$ and $\hP_{\by}$, respectively. The
rate functions, $R_X(\hP_{\bx})$ and $R_Y(\hP_{\by})$, are then given by 
$H(\hP_{\bu})$ and $H(\hP_{\bv})$, respectively, 
keeping in mind these correspondences between $\hP_{\bx}$ and $\hP_{\bu}$, and between
$\hP_{\by}$ and $\hP_{\bv}$.
In other words, for VRSW codes, $F(\hP_{\bu\bx})$ (resp.\ $G(\hP_{\bv\by})$) 
depends on $\hP_{\bx\bu}$ (resp.\ $\hP_{\by\bv}$) only via 
the marginals, $\hP_{\bx}$ and $\hP_{\bu}$ (resp.\ $\hP_{\by}$ and $\hP_{\bv}$).

The class of RBCs (\ref{genericW}) seems to offer much beyond the power of 
the ensembles of VRSW codes: it allows to shape
the empirical joint distribution, or the coordination, between 
$\bx$ and $\bu$ (resp.\ $\by$ and $\bv$). In fact, it is this 
coordination feature that makes (\ref{genericW})
significantly more general than the ordinary ensembles of S--W coding. 
It is deeper and more important
than the option of just selecting a non-uniform distribution under (\ref{genericW}). But what
is the motivation to shape such
an empirical correlation between each source sequence and its bin sequence?
The first obvious answer is that the ensemble of RBCs (\ref{genericW}) 
is more general than that of variable--rate S--W codes, and 
therefore, one might expect that the extra degrees
of freedom can potentially be used to gain improved performance 
in terms of the trade-off between the error exponent and the source coding exponent.
A somewhat deeper answer is best presented if we look at a 
simple example: suppose that $\by$ has already been decoded, and it now being used to
decode $\bx$ as well. Suppose also that $A$ assigns the uniform distribution over
a certain conditional type class of $\bu$ 
given $\bx$, say, all $\bu$--vectors that fall within Hamming distance 
1 away from $\bx$ (assuming, for simplicity,
that both $\bx$ and $\bu$ are binary vectors). Under such circumstances, 
the decoder knows that the only candidates for guessing $\bx$ based on $\bu$ and $\by$
must be within Hamming distance 1 from $\bu$. 
In other words, the number of incorrect source vectors, $\{\bx'\}$, that may 
confuse the decoder is significantly
reduced to lie within Hamming sphere of radius 1 around $\bu$. 
It appears then that, in a way, $\bu$ could serve as additional side information, beyond $\by$. 
Trivially, such a correlation between
$\bx$ and $\bu$ cannot improve the achievable compression rate of the source $X$, given by
the conditional entropy of $X$ given $Y$,
but at least on the face of it, one might speculate that 
the error exponent could be improved this way.

Our first result shows that, in spite of the above described 
idea of using $\bu$ for reducing the uncertainty in $\bx$, 
no coordination between $\bx$ and $\bu$ can possibly improve
the trade-off between error exponents and source coding exponents.
In other words, with the exception of a few pathological 
special cases (where the method of types is not applicable in a simple way),
the best trade-off between those exponents, that it is 
achievable with the class of RBCs (\ref{genericW}), 
can also be achieved using the smaller class of VRSW codes.
This means that the ensemble of VRSW codes is good in a stronger sense than that was known before.\footnote{In \cite{WM15}, it was shown that in
the simpler version of the S-W problem, where only one source is compressed and the other one serves as side information, the ensemble of
VRSW codes with an optimized rate function, are optimal at low rates, in the sense of meeting a converse bound. However, this applies only
at low enough rates, and it is based on the duality with channel coding and the sphere--packing bound. 
But this duality argument does not seem to be
applicable to the general S-W problem, where both sources are compressed.}

Nonetheless, the general RBCs of (\ref{genericW}) 
still have {\it raison d'\^etre} for at least two reasons.

\begin{enumerate}
\item
The first reason is that it covers the above--mentioned pathological special cases.
Consider, for example, a memoryless source $\{X_i\}$ with a 
uniform distribution over the alphabet $\{-3,-1,+1,+3\}$, and suppose that 
$y_i=|x_i|$, $i=1,2,\ldots,n$ ($\{x_i\}$ being realizations of $\{X_i\}$). 
Suppose further that $\by=(y_1,\ldots,y_n)$ is compressed to its entropy,
as is, which means that the RBC $B$ is the identity channel, 
that assigns probability one to $\bv=\by$, and that each $v_i$ is represented by one bit.
If the RBC $A$ assigns probability one to the sequence of signs, 
$u_i=\mbox{sgn}(x_i)$, $i=1,2,\ldots,n$ (which is also a special case 
of (\ref{genericW}), then both sources are perfectly recoverable at the decoder, 
as $x_i=u_iy_i$, and therefore the probability of error is strictly zero,
which means an infinite error exponent. The source coding exponent, i.e., 
the exponential rate of the probability that both code--lengths exceed a threshold
$n$, is also infinite, as there are exactly $n$ (uncompressed) bits for each source. 
On the other hand, for ordinary S--W codes, and even
for VRSW codes, the error exponent is finite even if one does 
not care about the source coding exponent. This is because there are 
in general many source sequences in each conditional type given the absolute values (all differing in permutations of signs) and some of them may be mapped into the
same bin.\footnote{Technically speaking, the reason that this example is exceptional is that the method of types fails, in this case, to assess the cardinality of the
set of incorrect typical source vectors, $\{\bx'\}$, that may confuse the decoder. While, in general, the cardinality of a typical set is assessed, exponentially
tightly, in terms of the corresponding empirical conditional entropy, in this particular case, this confusion set is empty, which is equivalent to replacing
the conditional empirical entropy by $-\infty$. More details can be found in the sequel.}

\item Maintaining coordination between $\bu$ and $\bx$, and between $\bv$ and $\by$, 
buys robustness in the following sense: imagine that due to some system failure, 
one of the compressed bit-streams, say, that of the $Y$-source, 
might not be available to the decoder, 
or that it might be severely corrupted. In this case, if a code drawn from 
the RBC (\ref{genericW}) is used, one can still reconstruct $\bu$ from 
its compressed bits, and by shaping the conditional distribution of $\bu$ given $\bx$,
one can control the distortion and keep it within a prescribed level $D$ away from $\bx$.
This is in sharp contrast to the situation with ordinary S--W codes, or even VRSW codes, 
		where the absence of side information could be totally catastrophic.\footnote{In the simpler version of the S--W problem,
		where only one source is compressed, and the other one serves as side information, this scenario is 
		a special case of a Wyner--Ziv setting where the side information may or may not be present, which was studied
		independently by Heegard and Berger \cite{HB85} and Kaspi \cite{Kaspi94}. Narrowing our model to the compression of
		one source only, our scheme may not be optimal in their sense, but it is considerably simpler.}
\end{enumerate}

The outline of the remaining part of the paper are as follows. In Section \ref{nc}, we establish notation conventions.
In Section \ref{pso}, we formalize the setting and spell out our objectives. In Section \ref{vrsw=opt},
we establish the asymptotic optimality of the ensemble of VRSW codes within the class of RBC's (\ref{genericW}).
In Section \ref{rbc}, we consider RBCs of the form (\ref{genericW}) and discuss the compromise between distortion,
error exponent, and source coding exponent. Finally, Section \ref{proofs} is devoted to proofs of our quantitative results.

\section{Notation Conventions}
\label{nc}

Throughout the paper, random variables will be denoted by capital
letters, specific values they may take will be denoted by the
corresponding lower case letters, and their alphabets
will be denoted by calligraphic letters. Random
vectors and their realizations will be denoted,
respectively, by capital letters and the corresponding lower case letters,
both in the bold face font. Their alphabets will be superscripted by their
dimensions. For example, the random vector $\bX=(X_1,\ldots,X_n)$, ($n$ --
positive integer) may take a specific vector value $\bx=(x_1,\ldots,x_n)$
in $\calX^n$, the $n$--th order Cartesian power of $\calX$, which is
the alphabet of each component of this vector.
Sources and channels will be denoted by capital letters, 
subscripted by the names of the relevant random variables/vectors and their
conditionings, if applicable, following the standard notation conventions,
e.g., $Q_X$, $P_{Y|X}$, and so on. When there is no room for ambiguity, these
subscripts will be omitted.
The probability of an event $\calE$ will be denoted by $\mbox{Pr}\{\calE\}$,
and the expectation
operator with respect to (w.r.t.) a probability distribution $P$ will be
denoted by
$\E_P\{\cdot\}$. Again, the subscript will be omitted if the underlying
probability distribution is clear from the context.
The entropy of a generic random variable $X$, with a distribution $Q$ on $\calX$, will be denoted by
$H_Q(X)$. Similar conventions will apply to other information measures. For example, 
if $(X,Y)$ is a pair of random variables, jointly distributed according to $Q_{XY}$ (or
just $Q$, for short), $H_Q(X,Y)$, $H_Q(X|Y)$ and $I_Q(X;Y)$ will denote the induced joint entropy, conditional entropy of $X$ given $Y$, and mutual information.
Similar notation rules will apply to larger sets of random variables (triplets, quadruplets, etc). 
The Kullback--Leibler divergence between two probability distributions,
$Q_{XY}$ and $P_{XY}$ is defined as
\begin{equation}
	D(Q_{XY}\|P_{XY})=\sum_{(x,y)\in\calX\times\calY}Q_{XY}(x,y)\log\frac{Q_{XY}(x,y)}{P_{XY}(x,y)},
\end{equation}
and the conditional weighted divergence is defined as
\begin{equation}
	D(Q_{Y|X}\|P_{Y|X}|Q_X)=\sum_{(x,y)\in\calX\times\calY}Q_{XY}(x,y)\log\frac{Q_{Y|X}(y|x)}{P_{Y|X}(y|x)},
\end{equation}
where logarithms, here and throughout the sequel, are understood to be taken to the base 2, unless specified otherwise.

For two
positive sequences $a_n$ and $b_n$, the notation $a_n\exe b_n$ will
stand for equality in the exponential scale, that is,
$\lim_{n\to\infty}\frac{1}{n}\log \frac{a_n}{b_n}=0$. Similarly,
$a_n\lexe b_n$ means that
$\limsup_{n\to\infty}\frac{1}{n}\log \frac{a_n}{b_n}\le 0$, and so on.
The indicator function
of an event $\calE$ will be denoted by $\calI\{E\}$. The notation $[x]_+$
will stand for $\max\{0,x\}$. The cardinality of a finite set, $\calA$, will be denoted by $|\calA|$.

The empirical distribution of a sequence $\bx\in\calX^n$, which will be
denoted by $\hat{P}_{\bx}$, is the vector of relative frequencies
$\hat{P}_{\bx}(x)$
of each symbol $x\in\calX$ in $\bx$.
The type class of $\bx\in\calX^n$, denoted $\calT(\bx)$, is the set of all
vectors $\bx'$
with $\hat{P}_{\bx'}=\hat{P}_{\bx}$. When we wish to emphasize the
dependence of the type class on the empirical distribution $\hat{P}$, we
will denote it by
$\calT(\hat{P})$. Information measures associated with empirical distributions
will be denoted with `hats' and will be subscripted by the sequences from
which they are induced. For example, the entropy associated with
$\hat{P}_{\bx}$, which is the empirical entropy of $\bx$, will be denoted by
$\hat{H}_{\bx}(X)$. Alternative notations
that will be used are $H(\hP_{\bx})$ and $\hH(\bx)$.
Similar conventions will apply to the joint empirical
distribution, the joint type class, the conditional empirical distributions
and the conditional type classes associated with pairs (and multiples) of
sequences of length $n$.
Accordingly, $\hP_{\bx\by}$ would be the joint empirical
distribution of $(\bx,\by)=\{(x_i,y_i)\}_{i=1}^n$,
$\calT(\bx,\by)$ or $\calT(\hP_{\bx\by})$ will denote
the joint type class of $(\bx,\by)$, $\calT(\bx|\by)$ will stand for
the conditional type class of $\bx$ given
$\by$, $\hH_{\bx\by}(X,Y)$ will designate the empirical joint entropy of $\bx$
and $\by$,
$\hH_{\bx\by}(X|Y)$ or $\hH(\bx|\by)$ will be the empirical conditional entropy,
$\hI_{\bx\by}(X;Y)$ will
denote empirical mutual information, and so on. The same notation rules apply also to
triplets and quadruplets of typical sequences, e.g., $\bu$, $\bv$, $\bx$ and $\by$.

\section{Problem Setting and Objectives}
\label{pso}

Consider a pair of sources defined by 
independent copies, $\{(X_i,Y_i),~i=1,2,\ldots\}$, of a pair of finite--alphabet 
random variables, $(X,Y)$, jointly distributed according
to a probability distribution, $P_{XY}$. The alphabets 
of $X$ and $Y$ are denoted by $\calX$ and $\calY$, respectively.
Both sequences are compressed separately and decoded jointly.

The encoder model is as follows. Each source vector, $\bx=(x_1,\ldots,x_n)$ (resp.\ 
$\by=(y_1,\ldots,y_n)$), 
is mapped into a bin, $f(\bx)=\bu=(u_1,\ldots,u_n)\in\calU^n$ 
(resp.\ $g(\by)=\bv=(v_1,\ldots,v_n)\in\calV^n$),
where $\calU$ (resp.\ $\calV$) is a finite alphabet, and then, 
the vector $\bu$ (resp.\ $\bv$) is compressed losslessly into a binary string. 
The decoder receives the two bit--streams. It first reconstructs
$\bu$ and $\bv$, and then uses them in order to estimate $(\bx,\by)$ using the MAP estimator,
\begin{equation}
	\label{MAPdecoder}
	(\hbx,\hby)=h[\bu,\bv]=\mbox{arg}\max_{\{(\bx,\by):~f(\bx)=\bu,
~g(\by)=\bv\}}P(\bx,\by),
\end{equation}
where 
\begin{equation}
	P(\bx,\by)=\prod_{i=1}^n P_{XY}(x_i,y_i).
\end{equation}
The mappings $f$ and $g$, henceforth referred to 
as the {\it bin--assignment mappings} (BAMs), are generated in 
the following manner. For each $\bx\in\calX^n$ (resp.\ $\by\in\calY^n$),
the corresponding bin $\bu$ (resp.\ $\bv$) is selected independently at 
random according to conditional distributions of the form
(\ref{genericW}), where $F(\cdot)$ and $G(\cdot)$ are arbitrary functionals 
of the respective empirical joint distributions, $\hP_{\bx\bu}$ and $\hP_{\by\bv}$, 
of the pair of vectors, $(\bx,\bu)$, and $(\by,\bv)$.
Once the BAMs have been selected, they are revealed to both encoder and decoder.

We are interested in the trade-off between two performance metrics. The first
is the average error probability with respect to (w.r.t.) the ensemble of BAMs,
\begin{equation}
	\bar{P}_{\mbox{\tiny err}}=\E\{\mbox{Pr}\{(\bX,\bY)\ne h[f(\bX),g(\bY)]\}=
	\sum_{\bu,\bv,\bx,\by}P(\bx,\by)A(\bu|\bx)B(\bv|\by)
\cdot\bar{P}_{\mbox{\tiny err}}(\bu,\bv,\bx,\by)
\end{equation}
where
\begin{equation}
	\bar{P}_{\mbox{\tiny err}}(\bu,\bv,\bx,\by)\dfn
\mbox{Pr}\left[\bigcup_{\{(\bx',\by')\ne(\bx,\by):~P(\bx',\by')\ge P(\bx,\by)\}}
\left\{f(\bx')=\bu,~g(\by')=\bv\right\}\right].
\end{equation}
Accordingly, we define the error exponent as
\begin{equation}
	\bE_{\mbox{\tiny err}}(A,B)=
\lim_{n\to\infty}\left[-\frac{\log \bar{P}_{\mbox{\tiny err}}}{n}\right],
\end{equation}
provided that the limit exists. We denote the error 
exponent as a functional of the RBCs $A$ and $B$, for
a reason that will become apparent shortly.

The second performance metric is the average 
{\it excess code-length probability} (a.k.a.\ the ``buffer overflow'' probability), defined
again w.r.t.\ the ensemble of BAMs (\ref{genericW}).
Given two prescribed threshold rate parameters, $\tR_X$ and $\tR_Y$, 
this probability is defined as
\begin{eqnarray}
\label{ecl}
	\bar{P}_{\mbox{\tiny ecl}}&=&\E\bigg[\min_{L_X,L_Y}\mbox{Pr}\{
L_X[f(\bX)]\ge n\tR_X,~L_Y[g(\bY)]\ge n\tR_Y\}\bigg]\nonumber\\
&=&\E\bigg\{\min_{L_X,L_Y}
\sum_{(\bx,\by):~L_X[f(\bx)]\ge n\tR_X,~L_Y[g(\by)]\ge n\tR_Y\}}P(\bx,\by)\bigg\},
\end{eqnarray}
where the expectation is over the ensemble of BAMs,
$L_X[\cdot]$ and $L_Y[\cdot]$ are length 
functions\footnote{Recall that a length function $L[\bu]=L[f(\bx)]$ of a source code is defined 
as the length (in bits) of the compressed representation of $\bu$. For a length function to correspond to a UD code, it must obey the Kraft inequality,
$\sum_{\bu}2^{-L(\bu)}\le 1$.}
of two uniquely decipherable (UD) lossless source codes, 
and $\mbox{Pr}\{\cdot\}$ is w.r.t.\ the randomness of $(\bX,\bY)$.
For given realization, $f$ and $g$, of the BAMs, the optimal length function, that achieves
the minimum in (\ref{ecl}) depends on the induced probability distributions of $\bu$ and $\bv$,
\begin{equation}
	P(\bu)=\sum_{\bx}P(\bx)\cdot\calI\{f(\bx)=\bu\},~~~
	P(\bv)=\sum_{\by}P(\by)\cdot\calI\{g(\by)=\bv\}.
\end{equation}
Since $f$ and $g$ are random, then so are $\{P(\bu)\}$ and $\{P(\bv)\}$.
The exact (asymptotic) analysis of $\bar{P}_{\mbox{\tiny ecl}}$ is non--trivial.
We therefore replace the definition of (\ref{ecl}) by a more convenient 
definition, as follows: instead of minimizing over $L_X$ and $L_Y$,
we take both of them to be the length function $L_*$ of the universal 
lossless code of memoryless sources, which represents $\bu$ (resp.\ $\bv$) in two parts:
the first is is the index of the type class of $\bu$ (resp.\ $\bv$) 
and the second is the index of $\bu$ (resp.\ $\bv$) within its type class. 
Since the number of types of
$n$--sequences is polynomial in $n$, the first part
requires $O(\log n)$ bits, whereas the second part 
requires $\log|\calT(\bu)|\approx nH(\hP_{\bu})$ (resp.\
$\log|\calT(\bv)|\approx nH(\hP_{\bv})$) bits. 
We now adopt the alternative definition of the excess code length probability,
\begin{equation}
	\tilde{P}_{\mbox{\tiny ecl}}=
\E\bigg[\mbox{Pr}\{L_*[f(\bX)]\ge n\tR_X,~L_*[g(\bY)]\ge n\tR_Y\}\bigg],
\end{equation}
which is obviously an upper bound on $\bar{P}_{\mbox{\tiny ecl}}$. 
The {\it excess code--length exponent} is then defined by
\begin{eqnarray}
	\bE_{\mbox{\tiny ecl}}(\tR_X,\tR_Y,A,B)&=&
\lim_{n\to\infty}\left[-\frac{\log \tilde{P}_{\mbox{\tiny ecl}}}{n}\right]\nonumber\\
	&=&\lim_{n\to\infty}\left[-\frac{\log 
\E[\mbox{Pr}\{H(\hP_{\bu})\ge \tR_X,~H(\hP_{\bv})\ge \tR_Y\}]}{n}\right].
\end{eqnarray}
For a given pair of threshold rates, $(\tR_X,\tR_Y)$, that satisfy 
$\tR_X > H(X|Y)$, $\tR_Y > H(Y|X)$, and $\tR_X+\tR_Y> H(X,Y)$, every 
pair $(A,B)$ of RBCs, achieves a certain error exponent, $\bE_{\mbox{\tiny err}}(A,B)$, and
a certain excess code--length exponent, $\bE_{\mbox{\tiny ecl}}(\tR_X,\tR_Y,A,B)$, 
i.e., a certain point 
$(\bE_{\mbox{\tiny err}}(A,B),\bE_{\mbox{\tiny ecl}}(\tR_X,\tR_Y,A,B))$ 
in the plane of the two exponents, and naturally, there is a certain tension
between them. We are interested in those RBC pairs $\{(A,B)\}$ that
achieve the best trade-off curve, that is, 
\begin{equation}
	\bE_{\mbox{\tiny err}}(E_{\mbox{\tiny ecl}},\tR_X,\tR_Y)
=\max\{\bE_{\mbox{\tiny err}}(A,B):~\bE_{\mbox{\tiny ecl}}(\tR_X,\tR_Y,A,B)\ge 
E_{\mbox{\tiny ecl}}\}, 
\end{equation}
where $E_{\mbox{\tiny ecl}}$ designates a prescribed excess 
code--length exponent level,\footnote{Note that
the excess code length exponent function is denoted 
with a bold face $\bE$, as opposed to the excess code length exponent level (which
is just a given non--negative real) is denoted in 
an ordinary font.} that varies between zero and infinity.

\section{Asymptotic Optimality of VRSW Codes}
\label{vrsw=opt}

Let the ensemble of S-W codes be given by eq.\ (\ref{genericW}), 
where the RBCs, $A$ and $B$, is 
defined with given functions, $F$ and $G$, respectively. 
Our first lemma provides general formulas for the error exponent 
and the source coding exponent.
\begin{lemma}
	\label{lemma1}
	Referring to the problem setting defined in Section \ref{pso}, 
	\begin{enumerate}
		\item The excess code--length exponent is given by
	\begin{eqnarray}
		\bE_{\mbox{\tiny ecl}}(\tR_X,\tR_Y,A,B)&=&
\min_{\{Q_{UVXY}:~H_Q(U)\ge \tR_X,~H_Q(V)\ge \tR_Y\}}
\{D(Q_{XY}\|P_{XY})+F(Q_{UX})+\nonumber\\
		& &G(Q_{VY})-H_Q(U|X)-H_Q(V|Y)\}.
	\end{eqnarray}
		\item Assuming that $(\bx,\by)$ cannot be uniquely determined from $(\bu,\bv)$,
\begin{equation}
\bE_{\mbox{\tiny err}}(A,B)=\min\{\bE_1(A),\bE_2(B),\bE_3(A,B)\},
\end{equation}
where
	\begin{eqnarray}
		\bE_1(A)&=&
\min_{Q_{UXY}}\{D(Q_{XY}\|P_{XY})+F(Q_{UX})-H_Q(U|X,Y)+\nonumber\\
		& &[F(Q_{UX})-H_Q(X|Y,U)]_+\}\\
		\bE_2(B)&=&
\min_{Q_{VXY}}\{D(Q_{XY}\|P_{XY})+G(Q_{VY})-H_Q(V|X,Y)+\nonumber\\
		& &[G(Q_{VY})-H_Q(Y|X,V)]_+\}\\
\bE_3(A,B)&=&
\min_{Q_{UVXY}}\{D(Q_{XY}\|P_{XY})+F(Q_{UX})+G(Q_{VY})-H_Q(U,V|X,Y)+\nonumber\\
		& &[F(Q_{UX})+G(Q_{VY})-H_Q(X,Y|U,V)]_+\}.
	\end{eqnarray}
	\end{enumerate}
\end{lemma}
In the second part of Lemma \ref{lemma1}, $\bE_1(A)$ is the error exponent associated with decoding errors in $\bx$ only,
$\bE_2(B)$ is the one associated with decoding errors in $\by$ only,
and $\bE_3(A,B)$ stands for simultaneous decoding errors in both $\bx$ and $\by$.
The second part of Lemma \ref{lemma1} 
generalizes the well--known formula for of the random coding 
error exponent of ordinary S-W codes, where
$F(Q_{UX})=H_Q(U)=R_X$ and $G(Q_{VY})=H_Q(V)=R_Y$. 
The minimization of the error exponent expression of 
$E_{\mbox{\tiny err}}(A,B)$ over $Q_{UVXY}$ 
under the constraint $H_Q(U)=R_X$, with $U$ being independent of $X$, 
causes $U$ and $V$ to be independent of $(X,Y)$, and of each other,
and then the error exponent formulas simplify to
(see, e.g., \cite{WM15}),
\begin{equation}
	\label{ordinarysw}
	 \bE_1(A)=\min_{Q_{XY}}\{D(Q_{XY}\|P_{XY})+[R_X-H_Q(X|Y)]_+\},
\end{equation}
and similar comments apply to $\bE_2(B)$ and $\bE_3(A,B)$.
It appears then that in the more general expression of $\bE_1(A,B)$,
$H_Q(X|Y)$ (in the square brackets) is replaced by $H_Q(X|Y,U)$. 
Likewise, in $\bE_2(A,B)$, $H_Q(Y|X)$ is replaced by $H_Q(Y|X,V)$ and in
$\bE_3(A,B)$, $H_Q(X,Y)$ is replaced by $H_Q(X,Y|U,V)$.
This is coherent with 
the insight that $U$ and $V$ play the role of additional side information, as described in the
Introduction. 

The next lemma gives a simple guideline for a good choice of the functions, $F$ and $G$.
\begin{lemma}
	\label{lemma2}
	Both $\bE_{\mbox{\tiny ecl}}(\tR_X,\tR_Y,A,B)$ and 
$\bE_{\mbox{\tiny err}}(A,B)$ are maximized by 
functions $F$ and $G$ with the following property: For every $Q_X$, there
	exists $Q_{U|X}^*$, which may depend 
on $Q_X$, such that
	\begin{equation}
		F(Q_{UX})=\left\{\begin{array}{ll}
				H_Q(U|X) & Q_{UX}=Q_X\times Q_{U|X}^*\\
				\infty & \mbox{elsewhere}\end{array}\right.
	\end{equation}
A similar argument applies to $G$ with $Q_X$, $Q_{UX}$ and $Q_{U|X}^*$, being replaced by
$Q_Y$, $Q_{VY}$, and $Q_{V|Y}^*$, respectively.
\end{lemma}
Lemma \ref{lemma2} tells us that the best RBCs put
all their mass on a single conditional type class of 
$\bu$ given $\bx$ (resp.\ $\bv$ given $\by$), 
wherein the distribution assigned is uniform,
i.e., 
\begin{equation}
	\label{Astar}
	A^*(\bu|\bx)=\left\{\begin{array}{ll}
		\frac{1}{|\calT(\bu|\bx)|} & \hP_{\bu\bx}=\hP_{\bx}\times Q_{U|X}^*\\
	0 & \mbox{elsewhere}\end{array}\right.
\end{equation}
and
\begin{equation}
	\label{Bstar}
	B^*(\bv|\by)=\left\{\begin{array}{ll}
		\frac{1}{|\calT(\bv|\by)|} & \hP_{\bv\by}=\hP_{\by}\times Q_{V|Y}^*\\
	0 & \mbox{elsewhere}\end{array}\right.
\end{equation}
This result is analogous to well known results about 
the optimality of ensembles of fixed composition with uniform distributions therein.

Combining Lemma \ref{lemma1} and Lemma \ref{lemma2}, 
we now obtain the following corollary concerning the 
expressions for the source coding exponent and the error exponent, where with slight abuse of notation,
we replace $A$ and $B$, in $\bE_{\mbox{\tiny ecl}}(\tR_X,\tR_Y,A,B)$ and $\bE_{\mbox{\tiny err}}(A,B)$,
by $Q_{U|X}^*$ and $Q_{V|Y}^*$, respectively.
\begin{corollary}
	\label{corollary}
\begin{equation}
	\bE_{\mbox{\tiny ecl}}(\tR_X,\tR_Y,Q_{U|X}^*,Q_{V|Y}^*)=\min_{\{Q_{UVXY}:~
H_Q(V)\ge \tR_Y\}}D(Q_{XY}\|P_{XY}),
\end{equation}
\begin{eqnarray}
	\bE_1(Q_{U|X}^*)&=&\min_{Q_{UXY}}\{D(Q_{XY}\|P_{XY})+H_Q(U|X)-
H_Q(U|X,Y)+\nonumber\\
	& &[H_Q(U|X)-H_Q(X|Y,U)]_+\},\\
	\bE_2(Q_{V|Y}^*)&=&\min_{Q_{VXY}}\{D(Q_{XY}\|P_{XY})+H_Q(V|Y)-
H_Q(V|X,Y)+\nonumber\\
	& &[H_Q(V|Y)-H_Q(Y|X,V)]_+\},\\
	\bE_3(Q_{U|X}^*,Q_{V|Y}^*)&=&\min_{Q_{UVXY}}\{D(Q_{XY}\|P_{XY})+H_Q(U|X)+H_Q(V|Y)-
H_Q(U,V|X,Y)+\nonumber\\
	& &[H_Q(U|X)+H_Q(V|Y)-H_Q(X,Y|U,V)]_+\},
\end{eqnarray}
	where all minimizations over $Q_{UVXY}$ are subject to the (additional) constraints, 
	$Q_{U|X}=Q_{U|X}^*$ (which is a function of $Q_X$), and
	$Q_{V|Y}=Q_{V|Y}^*$ (which is a function of $Q_Y$).
\end{corollary}
Note that for $\bE_{\mbox{\tiny ecl}}(\tR_X,\tR_Y,Q_{U|X}^*,Q_{V|Y}^*)$, 
the dependence on $Q_{U|X}^*$ and $Q_{V|Y}^*$ is only via the induced marginals, 
$Q_U^*$ and $Q_V^*$.
Corollary \ref{corollary} exhibits the seemingly non--trivial 
impact of generating dependencies between $X$ and $U$ and between $Y$ and $V$. 
On the one hand,
it helps as it reduces the subtracted terms, $H_Q(X|U,Y)$, $H_Q(Y|X,V)$ and $H_Q(X,Y|U,V)$ 
(as $U$ and $V$ serve as side informations), but on the other hand, it reduces also the
terms $H_Q(U|X)$ and $H_Q(V|Y)$, and it is not a priori obvious
what is the overall effect of such dependencies on the error exponent. 

The next result tells us, however, that 
for the best RBCs of the class of Lemma \ref{lemma2}, 
$Q_{U|X}^*=Q_U^*$ and $Q_{V|Y}^*=Q_V^*$, that is, $U$ is independent of $X$, and
$V$ is independent of $Y$, under $Q$.
\begin{theorem}
	\label{theorem1}
	For a given $Q_X$ (resp.\ $Q_Y$) and it is associated conditional distribution, $Q_{U|X}^*$ (resp.\ $Q_{V|Y}^*$),
	let $Q_U^*$ (resp.\ $Q_V^*$) denote the induced marginal. Then,
	\begin{eqnarray}
		\bE_{\mbox{\tiny ecl}}(\tR_X,\tR_Y,Q_U^*,Q_V^*)&=&\bE_{\mbox{\tiny ecl}}(\tR_X,\tR_Y,Q_{U|X}^*,Q_{V|Y}^*),\\
		\bE_{\mbox{\tiny err}}(Q_U^*,Q_V^*)&\ge&\bE_{\mbox{\tiny err}}(Q_{U|X}^*,Q_{V|Y}^*),\\
		\bE_{\mbox{\tiny err}}(Q_U^*,Q_V^*)&=&\min\{E_1(Q_{U}^*),E_2(Q_V^*),E_3(Q_U^*,Q_V^*)\}\\
		\bE_1(Q_U^*)&=&\min_{Q_{XY}}\{D(Q_{XY}\|P_{XY})+[H_{Q^*}(U)-H_Q(X|Y)]_+\}\\
		\bE_2(Q_V^*)&=&\min_{Q_{XY}}\{D(Q_{XY}\|P_{XY})+[H_{Q^*}(V)-H_Q(Y|X)]_+\}\\
		\bE_3(Q_U^*,Q_V^*)&=&\min_{Q_{XY}}\{D(Q_{XY}\|P_{XY})+[H_{Q^*}(U)+H_{Q^*}(V)-H_Q(X,Y)]_+\},
	\end{eqnarray}
	where $H_{Q^*}(U)$ and $H_{Q^*}(V)$ denote the entropies pertaining to $Q_U^*$ and $Q_V^*$, respectively.
\end{theorem}

The last part of Theorem \ref{theorem1} is identified with 
the error exponent of VRSW codes with rate functions $H_{Q^*}(U)$ and $H_{Q^*}(V)$ 
that depend on $Q_X$ and $Q_Y$, respectively. Indeed, as
explained in the Introduction, the ensemble of VRSW codes, 
with rate functions $R_X(\hP_{\bx})$ and $R_Y(\hP_{\by})$, 
corresponds to the special case of (\ref{genericW}), 
where $F(\hP_{\bu\bx})=H(\hP_{\bu})$ and
$G(\hP_{\by\bv})=H(\hP_{\bv})$, 
with $\hP_{\bu}$ and $\hP_{\bv}$ being chosen such that
$H(\hP_{\bu})=R_X(\hP_{\bx})$ and
$H(\hP_{\bv})=R_Y(\hP_{\by})$.\footnote{
This is a complete equivalence 
since the typical sequences of type $\hP_{\bu}$ 
can be mapped bijectively into binary sequences of length 
about $nH(\hP_{\bu})=nR_X(\hP_{\bx})$ and a similar comment applies to $\hP_{\bv}$. 
Note that the choice of the alphabets $\calU$ and $\calV$ is
completely immaterial in the case of VRSW codes. It is important only 
in the general case, where (\ref{genericW}) depends on the full joint empirical
distributions, $\hP_{\bu\bx}$ and $\hP_{\bv\by}$.}

\section{Using General Random Binning Channels for Robustness}
\label{rbc}

As mentioned in the Introduction, in spite of the conclusion of Theorem \ref{theorem1}, there are nevertheless at least two
motivations for using general RBC's that maintain some coordination between the source vectors and their associated bins.
The first is that it covers pathological, degenerate cases, as demonstrated in the Introduction. The other, and perhaps more
important motivation is that it allows more robustness to system failure: in the unfortunate event that the compressed
bit stream of one of the sources, say, $Y$, becomes unavailable to the decoder (for whatever reason), the other source, $X$, can 
still be reconstructed from its compressed bits within some distortion: if $Q_{U|X}^*$ is chosen such that
$\sum_{i=1}^nd(x_i,u_i)=n\sum_{x,u}Q_X(x)Q_{U|X}(u|x)d(x,u)\le nD$, for some distortion measure $d(\cdot,\cdot)$ and
distortion level $D$, then the reconstruction of $\bu$ can serve as an estimator of $\bx$.\footnote{Slightly more generally,
we could also use a certain function $q$ for the reconstruction, i.e.,
$\sum_{i=1}^nd(x_i,q(u_i))\le nD$, but for the sake of simplicity, we will let $q$ to be
the identity function.}
Obviously, this robustness comes
at the price of some degradation in the trade--off between the two other performance metrics: the error exponent and the
excess code--length exponent.\footnote{Moreover, the distortion constraints might incur penalties in terms of the achievable rates
altogether.} In other words, we now have a more general trade-off 
between three, rather than two, performance metrics:
the error exponent, the excess code--length exponent, and the
robustness to system failure, measured in terms of the distortions in the reconstruction of each source, in the event 
that the other source may not be available. More specifically, we would like to 
derive the following quantity for a given pair of threshold rates, $(\tR_X,\tR_Y)$:
\begin{eqnarray}
	\bE_{\mbox{\tiny err}}(E_{\mbox{\tiny ecl}},D_X,D_Y)&=&\max\bigg\{\bE_{\mbox{\tiny err}}(Q_{U|X},Q_{V|Y}):~
	\bE_{\mbox{\tiny ecl}}(\tR_X,\tR_Y,Q_{U|X},Q_{V|Y})\ge E_{\mbox{\tiny ecl}},\nonumber\\
	& &\sum_{x,u}Q_{UX}(u,x)d_X(x,u)\le D_X,~
	\sum_{y,v}Q_{VY}(v,y)d_Y(y,v)\le D_Y\bigg\},
\end{eqnarray}
where $d_X(\cdot,\cdot)$ and $d_Y(\cdot,\cdot)$ are distortion measures 
associated with the two sources, $D_X$ and $D_Y$ are the maximum allowable
distortion levels, and where we have removed the asterisks from $Q_{U|V}^*$ and $Q_{V|Y}^*$.
The requirement
$\bE_{\mbox{\tiny ecl}}(\tR_X,\tR_Y,Q_{U|X},Q_{V|Y}\ge E_{\mbox{\tiny ecl}}$ is equivalent to the requirement that
for every $Q_{XY}$ with $D(Q_{XY}\|P_{XY})\le E_{\mbox{\tiny ecl}}$, $Q_{U|X}$ and $Q_{V|Y}$ must be such that
$H_Q(U)\le\tR_X$ and $H_Q(V)\le\tR_Y$. In addition, the two distortion constraints must be satisfied.

In principle, given the earlier results presented above, 
the complete derivation of this quantity can be carried out in full generality, 
but it is evidently extremely complicated in
terms of the number of parameters to be optimized and the number of 
nested optimizations required. Therefore, in order to characterize some of the trade-offs under discussion, 
instead of the full optimization above,
we will confine ourselves
to the simpler case where the source $\bY$ is fully available at the decoder, and only $\bX$ must be decoded. 
This corresponds to the following formal choices in our setting: 
\begin{enumerate}
	\item $Q_{V|Y}$ is set to be the identity channel, i.e., $\bV=\bY$ with probability one.
	\item In view of item no.\ 1, we set $D_Y=0$.
	\item Since we have no requirements considering the excess code--length exponent of $\bY$, we formally 
		set $\tR_Y=0$. This degenerates
		the excess code--length event to be $\{L_X[f(\bX)]\ge n\tR_X\}$, or $\{H_Q(U)\ge\tR_X\}$ alone, without any
		requirement concerning the description length of $\bY$, since the event $\{L_Y[g(\bY)]\ge 0$ (or  $H_Q(V)\ge 0$)
		occurs with probability one, it is neutral in terms of intersections with other events.
\end{enumerate}
Since $E_2(Q_{V|Y})$ and $E_3(Q_{U|X},Q_{V|Y})$ are error exponents that are associated with decoding errors in $\bY$, they become irrelevant
in this case, and so, only $E_1(Q_{U|V})$ plays a role here.\footnote{Formally, we may consider $E_2(Q_{V|Y})$ 
and $E_3(Q_{U|X},Q_{V|Y})$ to be infinite.}

We have therefore simplified the problem as follows.
\begin{eqnarray}
	\bE_{\mbox{\tiny err}}(E_{\mbox{\tiny ecl}},\tR,D)&=&\max\bigg\{\bE_1(Q_{U|X}):~
	\bE_{\mbox{\tiny ecl}}(\tR,Q_{U|X})\ge E_{\mbox{\tiny ecl}},\nonumber\\
	& &\sum_{x,u}Q_{UX}(u,x)d(x,u)\le D\bigg\},
\end{eqnarray}
where we have omitted the subscript $X$ from $d_X$, $D_X$ and $\tR_X$, and we have redefined $\bE_{\mbox{\tiny ecl}}(\tR_X,Q_{U|V})$
to be
\begin{equation}
	\bE_{\mbox{\tiny ecl}}(\tR_X,Q_{U|X})=\min\{D(Q_X\|P_X):~H_Q(U)\ge \tR\}.
\end{equation}
Before we proceed, it would be convenient to represent $E_1(Q_{U|X})$ as follows:
\begin{eqnarray}
	E_1(Q_{U|X})&=&\min_{Q_X}\min_{Q_{Y|UX}}\{D(Q_{XY}\|P_{XY})+\nonumber\\
	& &H_Q(U|X)-H_Q(U|X,Y)+[H_Q(U|X)-H_Q(X|U,Y)]_+\}\nonumber\\
	&\dfn&\min_{Q_X}\zeta(Q_{UX}),
\end{eqnarray}
where we remind that $E_1(Q_{U|X})$ 
is defined for a given choice of $Q_{U|X}$ for every $Q_X$, and
where we have defined
\begin{equation}
	\zeta(Q_{UX})=\min_{Q_{Y|UX}}\{D(Q_{XY}\|P_{XY})+H_Q(U|X)-
H_Q(U|X,Y)+[H_Q(U|X)-H_Q(X|U,Y)]_+\}.
\end{equation}
In Subsection \ref{zeta}, we prove the following alternative expression for $\zeta(Q_{UX})$.
\begin{eqnarray}
	\label{zetaformula}
	\zeta(Q_{UX})&=&\inf_V\sup_{0\le\lambda\le 1}
\bigg\{\sum_{u,x}Q_{UX}(u,x)
\log\frac{[Q_{X}(x)]^{1+\lambda}}{P_X(x)[Q_U(u)]^\lambda}-\nonumber\\
	& &(1+\lambda)\sum_{u,x}Q_{UX}(u,x)\log Z(u,x,V,\lambda)\bigg\},
\end{eqnarray}
where
\begin{equation}
Z(u,x,V,\lambda)\dfn\sum_y
[P_{Y|X}(y|x)]^{1/(1+\lambda)}[V(y|u)]^{\lambda/(1+\lambda)},
\end{equation}
and where here $V=\{V(y|u),~u\in\calU,~y\in\calY\}$ (not to be confused with
the random variable $V$, which no longer plays a role here) is a 
matrix of conditional probabilities of $y$ given $u$, i.e.,
$V(y|u)\ge 0$ for all $(u,y)$ and $\sum_yV(y|u)=1$ for all $u$.

For $\bE_{\mbox{\tiny ecl}}(\tR,Q_{U|X})$ to exceed a prescribed level $E_{\mbox{\tiny ecl}}$,
every $Q_X$ with $D(Q_X\|P_X)\le E_{\mbox{\tiny ecl}}$, must correspond to $Q_{U|X}$ such that
the induced $Q_U$ satisfies ${\E}_Q\log[1/M(U)]\le \tR$ for
some probability distribution $M$ defined over $\calU$. 
Accordingly, we define
\begin{eqnarray}
	\calB(Q_X,M,\tR,D,E_{\mbox{\tiny ecl}})&=&
\{Q_{U|X}:~{\E}_Qd(X,U)\le D,~\mbox{and}~\nonumber\\
	& &D(Q_X\|P_X)\le E_{\mbox{\tiny ecl}}~\mbox{implies}~-{\E}_Q\log
	M(U)\le \tR\}.
\end{eqnarray}
Comment: we could have taken $M=Q$ and thereby replace the requirement
$-{\E}_Q\log M(U)\le \tR$ by $H_Q(U)\le \tR$, but this would make the problem
non--convex. Now, we have
\begin{eqnarray}
\bE_{\mbox{\tiny
	err}}(E_{\mbox{\tiny ecl}},\tR,D)&=&
	\inf_{Q_X}\sup_M\sup_{Q_{U|X}\in\calB(Q_X,M,\tR,D,E_{\mbox{\tiny ecl}})}
\zeta(Q_{UX})\nonumber\\
	&=&\min\{\hat{\bE}(\tR,E_{\mbox{\tiny ecl}},D),\tilde{\bE}(E_{\mbox{\tiny ecl}},D)\},
\end{eqnarray}
where
\begin{eqnarray}
	\hat{\bE}(\tR,E_{\mbox{\tiny ecl}},D)&=&\inf_{\{Q_X:~D(Q_X\|P_X)\le 
	E_{\mbox{\tiny ecl}}\}}\sup_M\sup_{\{Q_{U|X}:~-{\E}_Q\log
	M(U)\le \tR,~{\E}_Qd(X,U)\le D\}}\zeta(Q_{UX}),\\
	\tilde{\bE}(E_{\mbox{\tiny ecl}},D)&=&\inf_{\{Q_X:~D(Q_X\|P_X)> E_{\mbox{\tiny ecl}}\}}\sup_{\{Q_{U|X}:~\E _Qd(X,U)\le D\}}
\zeta(Q_{UX}).
\end{eqnarray}
Note that $\hat{\bE}(\tR,E_{\mbox{\tiny ecl}},D)$ 
is monotonically non-increasing in $E_{\mbox{\tiny ecl}}$ 
for fixed $(\tR,D)$, whereas
$\tilde{\bE}(E_{\mbox{\tiny ecl}},D)$ is monotonically 
non-decreasing, but the minimum between them,
$\bE_{\mbox{\tiny err}}(E_{\mbox{\tiny ecl}},\tR,D)$, is
non--increasing in $E_{\mbox{\tiny ecl}}$, because of the following consideration:
denoting 
\begin{equation}
K(Q_X)=\sup_M\sup_{\{Q_{U|X}:~-\E _Q\log M(U)\le \tR,~\E _Qd(X,U)\le D\}}\zeta(Q_{UX})
\end{equation}
and 
\begin{equation}
L(Q_X)=\sup_{\{Q_{U|X}:~\E _Qd(X,U)\le D\}}\zeta(Q_{UX}),
\end{equation}
we have
\begin{eqnarray}
\bE_{\mbox{\tiny
err}}(E_{\mbox{\tiny ecl}},\tR,D)&=&\inf_{Q_X}\bigg\{\calI[Q_X:~D(Q_X\|P_X)\le
E_{\mbox{\tiny ecl}}]\cdot K(Q_X)+\nonumber\\
	& &\calI[Q_X:~D(Q_X\|P_X)>E_{\mbox{\tiny ecl}}]\cdot L(Q_X)\bigg\}\\
&=&\inf_{Q_X}\left\{\calI[Q_X:~D(Q_X\|P_X)>
E_{\mbox{\tiny ecl}}]\cdot[L(Q_X)-K(Q_X)]+K(Q_X)\right\},
\end{eqnarray}
which is obviously monotonically non--increasing since $L(Q_X)\ge K(Q_X)$ and
$\calI[Q_X:~D(Q_X\|P_X)>E_{\mbox{\tiny ecl}}$ is non-increasing in $E_{\mbox{\tiny ecl}}$.
We now argue that $\hat{\bE}(\tR,E_{\mbox{\tiny ecl}},D)\le \tilde{\bE}(E_{\mbox{\tiny ecl}},D)$ 
and hence $\bE_{\mbox{\tiny
err}}(E_{\mbox{\tiny ecl}},\tR,D)=\hat{\bE}(\tR,E_{\mbox{\tiny ecl}},D)$ for all 
$E_{\mbox{\tiny ecl}}$, unless there is a range of $E_{\mbox{\tiny ecl}}$ where
$\tilde{\bE}(E_{\mbox{\tiny ecl}},D)$ is fixed, as a function of $E_{\mbox{\tiny ecl}}$ 
(i.e., the constraint is inactive). To see why this is true, suppose
conversely, that for some $\hat{E}_{\mbox{\tiny ecl}}>0$, we have $\hat{\bE}(\tR,\hat{E}_{\mbox{\tiny ecl}},D)>
\tilde{\bE}(\hat{E}_{\mbox{\tiny ecl}},D)$. Then for all $E_{\mbox{\tiny ecl}}<\hat{E}_{\mbox{\tiny ecl}}$, 
we have $\hat{\bE}(\tR,E_{\mbox{\tiny ecl}},D)\ge \hat{\bE}(\tR,\hat{E}_{\mbox{\tiny ecl}},D)
> \tilde{\bE}(\hat{E}_{\mbox{\tiny ecl}},D)\ge \tilde{\bE}(E_{\mbox{\tiny ecl}},D)$, which means that $\bE_{\mbox{\tiny
err}}(E_{\mbox{\tiny ecl}},\tR,D)=\tilde{\bE}(E_{\mbox{\tiny ecl}},D)$ for all $E_{\mbox{\tiny ecl}}\le\hat{E}_{\mbox{\tiny ecl}}$. 
Since $\tilde{\bE}(E_{\mbox{\tiny ecl}},D)$ is
monotonically non--deceasing as a function of $E_{\mbox{\tiny ecl}}$, 
then so is $\bE_{\mbox{\tiny err}}(E_{\mbox{\tiny ecl}},\tR,D)$ in the
range $E_{\mbox{\tiny ecl}}< \hat{E}_{\mbox{\tiny ecl}}$. But this is a contradiction to the non--increasing
monotonicity of $\bE_{\mbox{\tiny err}}(E_{\mbox{\tiny ecl}},\tR,D)$ unless $\tilde{\bE}(E_{\mbox{\tiny ecl}},D)$ is fixed
in the range $E_{\mbox{\tiny ecl}}<\hat{E}_{\mbox{\tiny ecl}}$, in which case it 
is equal to $\tilde{\bE}(0,D)$ throughout
this range. Thus, an alternative expression is
\begin{equation}
\bE_{\mbox{\tiny
err}}(E_{\mbox{\tiny ecl}},\tR,D)=
\min\{\hat{\bE}(\tR,E_{\mbox{\tiny ecl}},D),\tilde{\bE}(0,D)\}.
\end{equation}
We now have (see Subsection \ref{eteh}) 
the following lower bounds to $\hat{\bE}(\tR,E_{\mbox{\tiny ecl}},D)$ and $\tilde{\bE}(0,D)$.
\begin{eqnarray}
	\label{et}
	\tilde{\bE}(0,D)&\ge&
	\sup_{0\le\lambda\le 1}\inf_{\zeta\ge 0}\min_{C}\bigg[\zeta D-\nonumber\\
	& &(1+\lambda)\log\left\{
		\sum_x\min_u\left(2^{\zeta d(x,u)}\sum_y
	P(x,y)^{1/(1+\lambda)}C(u,y)^{\lambda/(1+\lambda)}\right)\right\}\bigg],
\end{eqnarray}
where $\{C(u,y)\}$ is an auxiliary probability distribution over $\calU\times\calY$, and
\begin{eqnarray}
	\label{eh}
	\hat{\bE}(\tR,E_{\mbox{\tiny ecl}},D)&\ge&
	\sup_{\theta\ge 0}\sup_{0\le\lambda\le 1}\inf_{\rho\ge 0}\inf_{\zeta\ge 0}
	\inf_V\sup_M\inf_W\bigg\{\rho \tR+\zeta D-\theta E_{\mbox{\tiny ecl}}-\nonumber\\
& &(1+\theta+\lambda)\log
\bigg(\sum_x\bigg[\frac{P_X^{1+\theta}(x)}{S(x,M,W,V,\rho,\zeta,\lambda)}\bigg]^{1/(1+\theta+\lambda)}\bigg)\bigg\}\nonumber\\
	&=&\sup_{\theta\ge 0}\sup_{0\le\lambda\le 1}\inf_{\rho\ge 0}\inf_{\zeta\ge 0}
	\bigg\{\rho \tR+\zeta D-\theta E_{\mbox{\tiny ecl}}-\nonumber\\
& &(1+\theta+\lambda)\sup_V\inf_M\sup_W\log
\bigg(\sum_x\bigg[\frac{P_X^{1+\theta}(x)}{S(x,M,W,V,\rho,\zeta,\lambda)}\bigg]^{1/(1+\theta+\lambda)}\bigg)\bigg\},
\end{eqnarray}
where
\begin{equation}
	S(x,M,W,V,\rho,\zeta,\lambda)\dfn\max_u\left\{\frac{M^\rho(u)}{W^\lambda(u)2^{\zeta d(x,u)}Z^{1+\lambda}(u,x,V,\lambda)}\right\},
\end{equation}
$M(\cdot)$ and $W(\cdot)$ being additional auxiliary probability distributions over $\calU$.

As we can see, in spite of the reduction to the simpler version of 
the S--W problem, the resulting exponent expression is still, by no means trivial, due to the
large number of nested optimizations, especially in eq.\ (\ref{eh}). 
Because of this excessive complexity of the expressions involved, we will 
not carry out a full investigation of the error exponent formula.
Instead, we will conclude this article by characterizing the behavior of the error exponent at
two important extreme special cases: $E_{\mbox{\tiny ecl}}=0$ and $E_{\mbox{\tiny ecl}}=\infty$. The rationale behind the focus on these
two extremes is two--fold: first, these two extremes are relatively simple, and secondly, they are relevant if one wishes
to contrast this with the behavior of the ensemble of fixed--rate S--W codes, 
where the excess code--length exponent is indeed always equal to either
zero or infinity, depending on whether $\tR$ is below or above the actual coding rate.
In both cases, we wish to characterize the
trade-off between the distortion $D$, between $U$ and $X$, and the threshold rate, $\tR$, such that the error exponent is still positive.
Since we are looking at the limit where the error exponent vanishes, we will assume that $\hat{\bE}(\tR,E_{\mbox{\tiny ecl}},D)$ dominates
the error exponent. 

In the case $E_{\mbox{\tiny ecl}}=0$, the maximizing conjugate parameter, $\theta$, tends to infinity.
Now, in this limit, we have that
\begin{eqnarray}
	& &\lim_{\theta\to\infty}(1+\theta+\lambda)\log\bigg(\sum_x
	\bigg[\frac{P_X^{1+\theta}(x)}{S(x,M,W,V,\rho,\zeta,\lambda)}\bigg]^{1/(1+\theta+\lambda)}\bigg)\nonumber\\
	&=&\lim_{\theta\to\infty}(1+\theta+\lambda)\log\bigg(\sum_x P_X(x)
	\bigg[\frac{P_X^{-\lambda}(x)}{S(x,M,W,V,\rho,\zeta,\lambda)}\bigg]^{1/(1+\theta+\lambda)}\bigg)\nonumber\\
	&=&\sum_xP_X(x)\log\bigg[\frac{P_X^{-\lambda}(x)}{S(x,M,W,V,\rho,\zeta,\lambda)}\bigg]\nonumber\\
	&=&\lambda H(X)-\sum_xP_X(x)\log S(x,M,W,V,\rho,\zeta,\lambda)\nonumber\\
	&\dfn& T_0(M,W,V,\rho,\zeta,\lambda).
\end{eqnarray}
Thus, the condition for a positive error exponent becomes
\begin{equation}
	\exists~0\le\lambda\le 1~\forall~\rho\ge 0,\zeta\ge 0,~\rho\tR+\zeta D \ge  \sup_V\inf_M\sup_WT_0(M,W,V,\rho,\zeta,\lambda),
\end{equation}
or, equivalently,
\begin{eqnarray}
	\tR&\ge&\inf_{0\le\lambda\le 1}\sup_{\rho\ge 0}\sup_{\zeta\ge 0}\frac{1}{\rho}
	\cdot[\sup_V\inf_M\sup_WT_0(M,W,V,\rho,\zeta,\lambda)-\zeta D]\nonumber\\
	&=&\inf_{0\le\lambda\le 1}\sup_{s\ge 0}\sup_{\zeta\ge 0}
	s[\sup_V\inf_M\sup_WT_0(M,W,V,1/s,\zeta,\lambda)-\zeta D].
\end{eqnarray}
In the case $E_{\mbox{\tiny ecl}}=\infty$, the maximizing conjugate parameter, $\theta$, must vanish. In this case, we get a similar
lower bound to the required threshold rate, $\tR$, except that $T_0$ is replaced by
\begin{eqnarray}
	T_\infty(M,W,V,\rho,\zeta,\lambda)&=&\lim_{\theta\to\infty}
	(1+\theta+\lambda)\log\bigg(\sum_x
        \bigg[\frac{P_X^{1+\theta}(x)}{S(x,M,W,V,\rho,\zeta,\lambda)}\bigg]^{1/(1+\theta+\lambda)}\bigg)\nonumber\\
	&=&(1+\lambda)\log\bigg(\sum_x
        \bigg[\frac{P_X(x)}{S(x,M,W,V,\rho,\zeta,\lambda)}\bigg]^{1/(1+\lambda)}\bigg).
\end{eqnarray}

\section{Proofs}
\label{proofs}

\subsection{Proof of Lemma \ref{lemma1}}

The first part of Lemma \ref{lemma1} is obtained by a simple application of the method of types \cite{CK81}.
\begin{eqnarray}
	\tilde{P}_{\mbox{\tiny ecl}}&=&\sum_{\bx,\by}\sum_{\{\bu:~H(\hP_{\bu})\ge 
\tR_X,~H(\hP_{\bv})\ge 
	\tR_Y\}}P(\bx,\by)A(\bu|\bx)B(\bv|\by)\nonumber\\
	&=&\sum_{\{\calT(\bu,\bv,\bx,\by):~H(\hP_{\bu})\ge tR_X,~H(\hP_{\bv})\ge tR_Y\}}
|\calT(\bu,\bv,\bx,\by,\bu,\bv)|\times\nonumber\\
	& &\exp_2\{-n[\hH_{\bx\by}(X,Y)+D(\hP_{\bx\by}\|P_{XY})+
F(\hP_{\bu\bx})+G(\hP_{\bv\by})]\}\nonumber\\
	&\exe&\max_{\{\calT(\bu,\bv,\bx,\by):~H(\hP_{\bu})\ge \tR_X,~H(\hP_{\bv})\ge\tR_Y\}}
	\exp_2\{n\hH_{\bu\bv\bx\by}(U,V,X,Y)\}\times\nonumber\\
	& &\exp_2\{-n[\hH_{\bx\by}(X,Y)+D(\hP_{\bx\by}\|P_{XY})+
F(\hP_{\bu\bx})+G(\hP_{\bv})]\}\nonumber\\
	&\exe&\max_{\{Q_{UVXY}:~H_Q(U)\ge \tR_X,~H_Q(V)\ge\tR_Y\}}\exp_2\{n[H_Q(X,Y,U,V)-H_Q(X,Y)-\nonumber\\
	& &D(Q_{XY}\|P_{XY})-F(Q_{UX})-G(Q_{VY})]\}\nonumber\\
	&=&\exp_2\bigg[-n\min_{\{Q_{UVXY}:~H_Q(U)\ge \tR_X,~H_Q(V)\ge\tR_Y\}}
\{D(Q_{XY}\|P_{XY})+\nonumber\\
	& &F(Q_{UX})+G(Q_{VY})-H_Q(U,V|X,Y)\}\bigg]\nonumber\\
	&=&\exp_2\bigg[-n\min_{\{Q_{XY},Q_{U|X},Q_{V|Y}:~H_Q(U)\ge\tR_X,~H_Q(V)\ge\tR_Y\}}
	\{D(Q_{XY}\|P_{XY})+F(Q_{UX})+\nonumber\\
	& &G(Q_{VY})-H_Q(U|X)-H_Q(V|Y)\}\bigg],
\end{eqnarray}
where the last step follows since the maximum of $H_Q(U,V|X,Y)$, for given $Q_{UX}$ 
and $Q_{VY}$, is achieved with
$H_Q(U,V|X,Y)=H_Q(U|X)+H_Q(V|Y)$.

Moving on the second part of Lemma \ref{lemma1}, 
we begin from an upper bound on the error probability.
Consider the sub-optimal decoder,
\begin{equation}
	(\tilde{\bx},\tilde{\by})=\mbox{arg}\min_{\{(\bx,\by):~f(\bx)=\bu,~g(\by)=\bv\}}m(\bu,\bv,\bx,\by),
\end{equation}
where the decoding metric, $m$, is defined by
\begin{equation}
	\label{univmetric}
	m(\bu,\bv,\bx,\by)=\max\{m_1(\bu,\bx,\by),m_2(\bv,\bx,\by),m_3(\bu,\bv,\bx,\by)\},
\end{equation}
where
\begin{eqnarray}
	m_1(\bu,\bv,\bx,\by)&=&\hH(\bx,\by|\bu,\bv)-F(\hP_{\bu\bx})-G(\hP_{\bv\by})\\
	m_2(\bu,\bx,\by)&=&\hH(\bx|\bu,\by)-F(\hP_{\bu\bx})\\
	m_3(\bv,\bx,\by)&=&\hH(\by|\bv,\bx)-G(\hP_{\bv\by}).
\end{eqnarray}
The error probability associated with this decoding metric is given by
\begin{eqnarray}
	\bar{P}_{\mbox{\tiny err}}&=&\sum_{\bu,\bv\bx,\by} 
P(\bx,\by)A(\bu|\bx)B(\bv|\by)\bar{P}_{\mbox{\tiny err}}(\bu,\bv,\bx,\by)\nonumber\\
	&=&\sum_{\bu,\bv,\bx,\by}P(\bx,\by)A(\bu|\bx)B(\bv|\by)\times\nonumber\\
	& &\mbox{Pr}\left[\bigcup_{\{(\bx',\by')\ne
(\bx,\by):~m(\bu,\bv,\bx',\by')\le
m(\bu,\bv,\bx,\by)\}}\left\{f(\bx')=f(\bx),~g(\by')=g(\by)\right\}\right]\nonumber\\
	&\dfn&\sum_{\bx,\by,\bu,\bv}P(\bx,\by)A(\bu|\bx)B(\bv|\by)\cdot \bar{P}_{\mbox{\tiny err}}(\bu,\bv,\bx,\by).
\end{eqnarray}
As for $\bar{P}_{\mbox{\tiny err}}(\bu,\bv,\bx,\by)$, we have
\begin{eqnarray}
\bar{P}_{\mbox{\tiny err}}(\bu,\bv,\bx,\by)&=&\mbox{Pr}\left[\bigcup_{\{(\bx',\by')\ne
(\bx,\by):~m(\bu,\bv,\bx',\by')\le
m(\bu,\bv,\bx,\by)\}}\left\{f(\bx')=f(\bx),~g(\by')=g(\by)\right\}\right]\nonumber\\
&\le&\mbox{Pr}\left[\bigcup_{\{\bx'\ne\bx,~\by'\ne\by:~m(\bu,\bv,\bx',\by')\le
m(\bu,\bv,\bx,\by)\}}\left\{
f(\bx')=f(\bx),~g(\by')=g(\by)\right\}\right]+\nonumber\\
& &\mbox{Pr}\left[\bigcup_{\{\bx'\ne\bx:~m(\bu,\bv,\bx',\by)\le m(\bu,\bv,\bx,\by)\}}
\left\{f(\bx')=f(\bx)\right\}\right]+\nonumber\\
	& &\mbox{Pr}\left[\bigcup_{\{\by'\ne\by:~m(\bu,\bv,\bx,\by')\le m(\bu,\bv,\bx,\by)\}}
\left\{g(\by')=g(\by)\right\}\right]\nonumber\\
&\dfn& \bar{P}_{\mbox{\tiny err},1}(\bu,\bv,\bx,\by)+
\bar{P}_{\mbox{\tiny err},2}(\bu,\bv,\bx,\by)+
\bar{P}_{\mbox{\tiny err},3}(\bu,\bv,\bx,\by).
\end{eqnarray}
Now,
\begin{eqnarray}
\bar{P}_{\mbox{\tiny
err},1}(\bu,\bv,\bx,\by)&=&\mbox{Pr}\left[\bigcup_{\{\tbx\ne\bx,~\tby\ne\by:~m(\bu,\bv,\tbx,\tby)\le
m(\bu,\bv,\bx,\by)\}}\left\{
f(\tbx)=f(\bx),~g(\tby)=g(\by)\right\}\right]\nonumber\\
&\le&\min\left\{1,A(\bu|\tbx)B(\bv|\tby)\bigg|\left\{(\tbx,\tby):~m(\bu,\bv,\tbx,\tby)\le
m(\bu,\bv,\bx,\by)\right\}\bigg|\right\}\nonumber\\
	&=&\min\bigg\{1,\exp_2\{-n[F(\hP_{\bu\tbx})+G(\hP_{\bv\tby})]\}\times\nonumber\\
	& &\bigg|
\left\{(\tbx,\tby):~m(\bu,\bv,\tbx,\tby)\le
	m(\bu,\bv,\bx,\by)\right\}\bigg|\bigg\}\nonumber\\
&\le&\min\bigg\{1,\exp_2\{-n[F(\hP_{\bu\tbx})+G(\hP_{\bv\tby})]\}\cdot
\sum_{\{(\calT(\tbx,\tby|\bu,\bv):~
~m(\bu,\bv,\tbx,\tby)\le
m(\bu,\bv,\bx,\by)\}}\nonumber\\
& &|\calT(\tbx,\tby|\bu,\bv)|
\bigg\}\nonumber\\
&\le&\min\bigg\{1,
\sum_{\{\calT(\tbx,\tby|\bu,\bv):~
m(\bu,\bv,\tbx,\tby)\le
m(\bu,\bv,\bx,\by)\}}
	\exp_2\{n[\hH(\tbx,\tby|\bu,\bv)-\nonumber\\
	& &F(\hP_{\bu\tbx})-G(\hP_{\bv\tby})]\}\bigg\}
	\nonumber\\
&=&\min\bigg\{1,
\sum_{\{\calT(\tbx,\tby|\bu,\bv):~
m(\bu,\bv,\tbx,\tby)\le
m(\bu,\bv,\bx,\by)\}}
	\exp_2[nm_1(\bu,\bv,\tbx,\tby)]\bigg\}.
\end{eqnarray}
In the same fashion, we obtain (see also \cite{asyncsw})
\begin{eqnarray}
	\bar{P}_{\mbox{\tiny err},2}&\lexe&\min\bigg\{1,
\sum_{\{\calT(\tbx|\bu,\by):~
m(\bu,\bv,\tbx,\by)\le
m(\bu,\bv,\bx,\by)\}}
        \exp_2[nm_2(\bu,\bv,\tbx,\by)]\bigg\}\\
	\bar{P}_{\mbox{\tiny err},3}&\lexe&\min\bigg\{1,
\sum_{\{\calT(\tby|\bv,\bx):~
m(\bu,\bv,\tbx,\by)\le
m(\bu,\bv,\bx,\by)\}}
        \exp_2[nm_3(\bu,\bv,\bx,\tby)]\bigg\},
\end{eqnarray}
and so,
\begin{eqnarray}
	\bar{P}_{\mbox{\tiny err}}&\lexe&\min\bigg\{1,\max_{\{\hP_{\tbx\tby|\bu\bv}:~m(\bu,\bv,\tbx,\tby)\le m(\bu,\bv,\bx,\by)\}}
	\{\exp_2[nm_1(\bu,\bv,\tbx,\tby)]+\nonumber\\
	& &\exp_2[nm_2(\bu,\tbx,\tby)]+\exp_2[nm_3(\bv,\tbx,\tby)]\}\bigg\}\\
	&\exe&\min\bigg\{1,\max_{\{\hP_{\tbx\tby|\bu\bv}:~m(\bu,\bv,\tbx,\tby)\le m(\bu,\bv,\bx,\by)\}}
	\exp_2[nm(\bu,\bv,\tbx,\tby)]\}\bigg\}\\
	&=&\min\bigg\{1,
	\exp_2[nm(\bu,\bv,\bx,\by)]\}\bigg\}.
\end{eqnarray}
Using the method of types, the resulting error exponent is
\begin{eqnarray}
	E_{\mbox{\tiny err}}(A,B)&\ge&\min_{Q_{UVXY}}\bigg\{D(Q_{XY}\|P_{XY})+
	F(Q_{UX})+G(Q_{VY})-H_Q(U,V|X,Y)+\nonumber\\
		& &[\min\{F(Q_{UX})-H_Q(X|U,Y),G(Q_{VY})-H_Q(Y|V,X),\nonumber\\
		& &F(Q_{UX})+G(Q_{VY})-H_Q(X,Y|U,V)\}]_+\bigg\}\nonumber\\
	&=&\min_{Q_{UVXY}}\bigg\{D(Q_{XY}\|P_{XY})+
	F(Q_{UX})+G(Q_{VY})-H_Q(U,V|X,Y)+\nonumber\\
	& &\min\{[F(Q_{UX})-H_Q(X|U,Y)]_+,[G(Q_{VY})-H_Q(Y|V,X)]_+,\nonumber\\
	& &[F(Q_{UX})+G(Q_{VY})-H_Q(X,Y|U,V)]_+\}\bigg\}\nonumber\\
	&=&\min\{\bE_1(A),\bE_2(B),\bE_3(A,B)\},
\end{eqnarray}
where the inequality is due to the fact that we have analyzed a sub--optimal decoder,
whereas $\bE_{\mbox{\tiny err}}(A,B)$ is defined for the optimal, MAP decoder (\ref{MAPdecoder}).
A matching lower bound, of the same exponential order, 
is obtained by analyzing the error probability for a given $(\bu,\bv,\bx,\by)$
under the MAP decoder, where the inner summation over $\{(\bx',\by')\}$ 
is confined to the same conditional type
as that of $(\bx,\by)$ given $(\bu,\bv)$ (where 
$P(\bx',\by')=P(\bx,\by)$ and $A(\bu|\bx')=A(\bu|\bx)$), see, \cite{asyncsw} where the
same idea was used as well. 
Since the sub-optimal decoding metric (\ref{univmetric}) cannot be better 
than the MAP decoding metric, a simple sandwich argument
yields that the last inequality is, in fact, an equality:
\begin{equation}
	\bE_{\mbox{\tiny err}}(A,B)=\min\{\bE_1(A),\bE_2(B),\bE_3(A,B)\}.
\end{equation}
This completes the proof of Lemma \ref{lemma1}.

\subsection{Proof of Lemma \ref{lemma2}}

First, observe that for both $E_{\mbox{\tiny err}}(A,B)$ and 
$E_{\mbox{\tiny ecl}}(\tR_X,\tR_Y,A,B)$
to improve, the larger we can make $F$ and $G$ -- the better. 
But since it is associated with a conditional distribution, there is an inherent limitation,
which stems from the following simple derivation:
\begin{equation}
	1=\sum_{\bu}A(\bu|\bx)=\sum_{\{\calT(\bu|\bx)\}}|\calT(\bu|\bx)|\cdot 
2^{-nF(\hP_{\bu\bx})}\exe\max_{Q_{U|X}} 2^{n[H_Q(U|X)-F(Q_{UX})]},
\end{equation}
or equivalently, $\max_{Q_{U|X}}[H_Q(U|X)-F(Q_{UX})]=0$. 
In other words, for each $Q_X$, the following holds true: for every $Q_{U|X}$,
$F(Q_{UX})\ge H_Q(U|X)$ and the 
inequality must be met with equality for at least one conditional distribution, $Q_{U|X}$.
The best one can in maximizing $F(Q_{UX})$ 
(and hence also both exponents) is therefore to take it to infinity for all $Q_{U|X}$, except one
conditional distribution, $Q_{U|X}$, 
which we denote by $Q_{U|X}^*$, and which may depend on $Q_X$.
The same line of thought applies, of course, to $G$.

\subsection{Proof of Theorem \ref{theorem1}}

The first part, 
$\bE_{\mbox{\tiny ecl}}(\tR_X,\tR_Y,Q_U^*,Q_V^*)=
\bE_{\mbox{\tiny ecl}}(\tR_X,\tR_Y,Q_{U|X}^*,Q_{V|Y}^*)$, 
is obvious since the dependence of 
$\bE_{\mbox{\tiny ecl}}(\tR_X,\tR_Y,Q_{U|X}^*,Q_{V|Y}^*)$ on $Q_{U|X}^*$ and $Q_{V|Y}^*$ 
is only via its unconditional marginals, $Q_U^*$ and $Q_V^*$, 
anyway. As for the second part,
consider the following chain of inequalities, for a given $Q_X$ and $Q_{U|X}^*$:
\begin{eqnarray}
	& &\min_{Q_{Y|UX}}\{D(Q_{XY}\|P_{XY})+H_Q(U|X)-H_Q(U|X,Y)+[H_Q(U|X)-H_Q(X|Y,U)]_+\}\\
	&\le&\min_{Q_{Y|UX}=Q_{Y|X}}\{D(Q_{XY}\|P_{XY})+H_Q(U|X)-H_Q(U|X,Y)+\nonumber\\
	& &[H_Q(U|X)-H_Q(X|Y,U)]_+\}\\
	&=&\min_{Q_{Y|UX}=Q_{Y|X}}\{D(Q_{XY}\|P_{XY})+H_Q(U|X)-H_Q(U|X)+\nonumber\\
	& &[H_Q(U|X)-H_Q(X|Y)+I_Q(X;U|Y)]_+\}\\
	&=&\min_{Q_{Y|UX}=Q_{Y|X}}\{D(Q_{XY}\|P_{XY})+[H_Q(U|X)-H_Q(X|Y)+H_Q(U|Y)-H_Q(U|X,Y)]_+\}\\
	&=&\min_{Q_{Y|UX}=Q_{Y|X}}\{D(Q_{XY}\|P_{XY})+[H_Q(U|X)-H_Q(X|Y)+H_Q(U|Y)-H_Q(U|X)]_+\}\\
	&=&\min_{Q_{Y|UX}=Q_{Y|X}}\{D(Q_{XY}\|P_{XY})+[H_Q(U|Y)-H_Q(X|Y)]_+\}\\
	&\le&\min_{Q_{Y|UX}=Q_{Y|X}}\{D(Q_{XY}\|P_{XY})+[H_Q(U)-H_Q(X|Y)]_+\}\\
	&\le&\min_{Q_{Y|X}}\{D(Q_{XY}\|P_{XY})+[H_Q(U)-H_Q(X|Y)]_+\}.
\end{eqnarray}
Minimizing the first and the last expressions in the above chain over $Q_X$, while holding $Q_{U|X}=Q_{U|X}^*$ for each $Q_X$,
we get
\begin{equation}
	\label{almostdone1}
	\bE_1(Q_{U|X}^*)\le\min_{Q_{XY}}\{D(Q_{XY}\|P_{XY})+[H_{Q^*}(U)-H_Q(X|Y)]_+\},
\end{equation}
Replacing $U$ by $V$ and interchanging the roles of $X$ and $Y$, we similarly obtain
\begin{equation}
	\label{almostdone2}
	\bE_2(Q_{V|Y}^*)\le\min_{Q_{XY}}\{D(Q_{XY}\|P_{XY})+[H_{Q^*}(V)-H_Q(Y|X)]_+\}.
\end{equation}
Likewise, let us define $\calS\dfn\{Q_{UVXY}:~Q_{U|X}=Q_{U|X}^*,~Q_{V|Y}=Q_{V|Y}^*\}$ and
let $\calS'=\calS\cap\{Q_{UVXY}:~ Q_{UVXY}=Q_{XY}\times Q_{U|X}\times Q_{V|Y}\}$. Then,
\begin{eqnarray}
	\label{almostdone3}
	\bE_3(A,B)&=&\min_{Q_{UVXY}\in\calS}\{D(Q_{XY}\|P_{XY})+H_Q(U|X)+H_Q(V|Y)-H_Q(U,V|X,Y)+\nonumber\\
	& &[H_Q(U|X)+H_Q(V|Y)-H_Q(X,Y|U,V)]_+\}\\
	&\le&\min_{Q_{UVXY}\in\calS'}\{D(Q_{XY}\|P_{XY})+H_Q(U|X)+H_Q(V|Y)-H_Q(U,V|X,Y)+\nonumber\\
	& &[H_Q(U|X)+H_Q(V|Y)-H_Q(X,Y)+I_Q(X,Y;U,V)]_+\}\\
	&=&\min_{Q_{UVXY}\in\calS'}\{D(Q_{XY}\|P_{XY})+H_Q(U|X)+H_Q(V|Y)-H_Q(U|X)-H_Q(V|Y)+\nonumber\\
	& &[H_Q(U|X)+H_Q(V|Y)-H_Q(X,Y)+H_Q(U,V)-H_Q(U,V|X,Y)]_+\}\\
	&=&\min_{Q_{UVXY}\in\calS'}\{D(Q_{XY}\|P_{XY})+\nonumber\\
	& &[H_Q(U|X)+H_Q(V|Y)-H_Q(X,Y)+H_Q(U,V)-H_Q(U|X)-H_Q(V|Y)]_+\}\\
	&=&\min_{Q_{UVXY}\in\calS'}\{D(Q_{XY}\|P_{XY})+[H_Q(U,V)-H_Q(X,Y)]_+\}\\
	&\le&\min_{Q_{UVXY}\in\calS'}\{D(Q_{XY}\|P_{XY})+[H_Q(U)+H_Q(V)-H_Q(X,Y)]_+\}\\
	&=&\min_{Q_{XY}}\{D(Q_{XY}\|P_{XY})+[H_Q(U)+H_Q(V)-H_Q(X,Y)]_+\},
\end{eqnarray}
and the proof of Theorem \ref{theorem1} (including the last part) 
is now completed 
once one observes that the right--hand sides of (\ref{almostdone1}), (\ref{almostdone2}) 
and (\ref{almostdone3}) are exactly the expressions obtained for
$\bE_{\mbox{\tiny err}}(Q_{U}^*,Q_V^*)$. Indeed, for $Q_{U|X}^*=Q_U^*$ and $Q_{V|Y}^*=Q_V^*$, the minimizing $Q_{UVXY}$ 
in Corollary \ref{corollary} is such that
$(X,Y)$, $U$ and $V$ are all independent, which 
yields $H_Q(U|X,Y)=H_Q(U|X)=H_Q(U)$, $H_Q(V|X,Y)=H_Q(V|Y)=H_Q(V)$,
$H_Q(X,Y|U,V)=H_Q(X,Y)$, $H_Q(X|Y,U)=H_Q(X|Y)$, and $H_Q(Y|X,V)=H_Q(Y|X)$.

\subsection{Proof of Eq.\ (\ref{zetaformula})}
\label{zeta}

\begin{eqnarray}
	\zeta(Q_{UX})&=&\inf_{Q_{Y|UX}}\{D(Q_{XY}\|P_{XY})+H_Q(U|X)-H_Q(U|X,Y)+\nonumber\\
	& &[H_Q(U|X)-H_Q(X|U,Y)]_+\}\nonumber\\
	&=&\inf_{Q_{Y|UX}}\{D(Q_{XY}\|P_{XY})+H_Q(Y|X)-H_Q(Y|U,X)+\nonumber\\
	& &[H_Q(U|X)-H_Q(X|U)+H_Q(Y|U)-H_Q(Y|U,X]_+\}\nonumber\\
	&=&\inf_{Q_{Y|UX}}\{D(Q_{XY}\|P_{XY})+H_Q(Y|X)-H_Q(Y|U,X)+\nonumber\\
	& &[H_Q(U)-H_Q(X)+H_Q(Y|U)-H_Q(Y|U,X)]_+\}\nonumber\\
&=&\inf_{Q_{Y|UX}}\sup_{0\le\lambda\le
1}\{D(Q_{XY}\|P_{XY})+H_Q(Y|X)-(1+\lambda)H_Q(Y|U,X)+\nonumber\\
	& &\lambda[H_Q(U)-H_Q(X)+H_Q(Y|U)]\nonumber\\
&=&\inf_{Q_{Y|UX}}\sup_{0\le\lambda\le
1}\inf_V\sum_{u,x,y}Q_{UXY}(u,x,y)\bigg[\log\frac{Q_{XY}(x,y)}{P_{XY}(x,y)}-\log
Q_{Y|X}(y|x)+\nonumber\\
	& &(1+\lambda)\log Q_{Y|UX}(y|u,x)-
\lambda\log Q_U(u)+\lambda\log
Q_X(x)-\lambda\log V(y|u)\bigg]\nonumber\\
&=&\inf_{Q_{Y|UX}}\sup_{0\le\lambda\le
1}\inf_V\sum_{u,x,y}Q_{UXY}(u,x,y)\log\frac{Q_{XY}(x,y)
[Q_{Y|UX}(y|u,x)]^{1+\lambda}[Q_X(x)]^\lambda}{P_{XY}(x,y)Q_{Y|X}(y|x)[Q_U(u)]^\lambda
[V(y|u)]^\lambda}\nonumber\\
&=&\inf_{Q_{Y|UX}}\sup_{0\le\lambda\le
1}\inf_V\sum_{u,x,y}Q_{UXY}(u,x,y)\log\frac{Q_{X}(x)
[Q_{Y|UX}(y|u,x)]^{1+\lambda}[Q_X(x)]^\lambda}{P_{XY}(x,y)[Q_U(u)]^\lambda
[V(y|u)]^\lambda}\nonumber\\
&=&\inf_{Q_{Y|UX}}\sup_{0\le\lambda\le
1}\inf_V\sum_{u,x,y}Q_{UXY}(u,x,y)\log\frac{[Q_{X}(x)]^{1+\lambda}
[Q_{Y|UX}(y|u,x)]^{1+\lambda}}{P_{XY}(x,y)[Q_U(u)]^\lambda
[V(y|u)]^\lambda}\nonumber\\
&=&\inf_{Q_{Y|UX}}\sup_{0\le\lambda\le
1}\inf_V\bigg\{\sum_{u,x}Q_{UX}(u,x)
\log\frac{[Q_{X}(x)]^{1+\lambda}}{P_X(x)[Q_U(u)]^\lambda}+\nonumber\\
& &\sum_{u,x}Q_{UX}(u,x)\sum_yQ_{Y|UX}(y|u,x)\log
\frac{[Q_{Y|UX}(y|u,x)]^{1+\lambda}}{P_{Y|X}(y|x)[V(y|u)]^\lambda}\bigg\}\nonumber\\
&=&\inf_{Q_{Y|UX}}\inf_V\sup_{0\le\lambda\le
1}\bigg\{\sum_{u,x}Q_{UX}(u,x)\log\frac{[Q_{X}(x)]^{1+\lambda}}
{P_X(x)[Q_U(u)]^\lambda}+\nonumber\\
& &(1+\lambda)\sum_{u,x}Q_{UX}(u,x)\sum_yQ_{Y|UX}(y|u,x)\log
\frac{Q_{Y|UX}(y|u,x)}{[P_{Y|X}(y|x)]^{1/(1+\lambda)}
[V(y|u)]^{\lambda/(1+\lambda)}}\bigg\}\nonumber\\
&=&\inf_V\inf_{Q_{Y|UX}}\sup_{0\le\lambda\le
1}\bigg\{\sum_{u,x}Q_{UX}(u,x)\log\frac{[Q_{X}(x)]^{1+\lambda}}
{P_X(x)[Q_U(u)]^\lambda}+\nonumber\\
& &(1+\lambda)\sum_{u,x}Q_{UX}(u,x)\sum_yQ_{Y|UX}(y|x,u)\log
\frac{Q_{Y|UX}(y|x,u)}{[P_{Y|X}(y|x)]^{1/(1+\lambda)}
[V(y|u)]^{\lambda/(1+\lambda)}}\bigg\}\nonumber\\
&=&\inf_V\sup_{0\le\lambda\le 1}\inf_{Q_{Y|UX}}
\bigg\{\sum_{u,x}Q_{UX}(u,x)\log\frac{[Q_{X}(x)]^{1+\lambda}}{P_X(x)[Q_U(u)]^\lambda}+\nonumber\\
& &(1+\lambda)\sum_{u,x}Q_{UX}(u,x)\sum_yQ_{Y|UX}(y|u,x)\log
\frac{Q_{Y|UX}(y|u,x)}{[P_{Y|X}(y|x)]^{1/(1+\lambda)}
[V(y|u)]^{\lambda/(1+\lambda)}}\bigg\}\nonumber\\
&=&\inf_V\sup_{0\le\lambda\le 1}
\bigg\{\sum_{u,x}Q_{UX}(u,x)\log\frac{[Q_{X}(x)]^{1+\lambda}}{P_X(x)[Q_U(u)]^\lambda}-\nonumber\\
& &(1+\lambda)\sum_{u,x}Q_{UX}(u,x)\log\bigg(\sum_y
[P_{Y|X}(y|x)]^{1/(1+\lambda)}[V(y|u)]^{\lambda/(1+\lambda)}\bigg)\bigg\}.
\end{eqnarray}

\subsection{Proof of Eqs.\ (\ref{et}) and (\ref{eh})}
\label{eteh}

As for eq.\ (\ref{et}), we have:
\begin{eqnarray}
	\tilde{\bE}(E_{\mbox{\tiny ecl}},D)&=&\inf_{Q_X}\sup_{\{Q_{U|X}:~{\E}_Qd(X,U)\le D\}}\zeta(Q_{UX})\\
	&=&\inf_{Q_X}\sup_{Q_{U|X}}\inf_{\zeta\ge 0}\inf_V\sup_{0\le\lambda\le 1}
	\bigg\{\sum_{u,x}Q_{UX}(u,x)\log\frac{[Q_{X}(x)]^{1+\lambda}}{P_X(x)[Q_U(u)]^\lambda}+\zeta[D-{\E}_Qd(X,U)]-\nonumber\\
& &(1+\lambda)\sum_{u,x}Q_{UX}(u,x)\log Z(u,x,V,\lambda)\bigg\}\\
	&=&\inf_{Q_X}\sup_{Q_{U|X}}\inf_{\zeta\ge 0}\inf_V\sup_{0\le\lambda\le 1}\min_W
\bigg\{\zeta D+D(Q_X\|P_X)-\lambda H_Q(X)-\nonumber\\
	& &\lambda{\E}_Q\log W(U)-\zeta{\E}_Qd(X,U)-\nonumber\\
& &(1+\lambda)\sum_{u,x}Q_{UX}(u,x)\log Z(u,x,V,\lambda)\bigg\}\\
	&=&\inf_{Q_X}\sup_{Q_{U|X}}\inf_{\zeta\ge 0}\inf_V\sup_{0\le\lambda\le 1}\min_W
\bigg\{\zeta D+D(Q_X\|P_X)-\lambda H_Q(X)-\nonumber\\
	& &\sum_xQ_X(x)\sum_uQ_{U|X}(u|x)\log[W^\lambda(u)2^{\zeta d(x,u)}
Z^{1+\lambda}(u,x,V,\lambda)]\bigg\}\\
	&=&\inf_{Q_X}\sup_{Q_{U|X}}\inf_{\zeta\ge 0}\sup_{0\le\lambda\le 1}\min_{V,W}
\bigg\{\zeta D+D(Q_X\|P_X)-\lambda H_Q(X)-\nonumber\\
	& &\sum_xQ_X(x)\sum_uQ_{U|X}(u|x)\log[W^\lambda(u)2^{\zeta d(x,u)}
Z^{1+\lambda}(u,x,V,\lambda)]\bigg\}\\
	&=&\inf_{Q_X}\sup_{0\le\lambda\le 1}\sup_{Q_{U|X}}\inf_{\zeta\ge 0}\min_{V,W}
\bigg\{\zeta D+D(Q_X\|P_X)-\lambda H_Q(X)-\nonumber\\
	& &\sum_xQ_X(x)\sum_uQ_{U|X}(u|x)\log[W^\lambda(u)2^{\zeta d(x,u)}
Z^{1+\lambda}(u,x,V,\lambda)]\bigg\}\\
	&=&\inf_{Q_X}\sup_{0\le\lambda\le 1}\inf_{\zeta\ge 0}\min_{V,W}\sup_{Q_{U|X}}
\bigg\{\zeta D+D(Q_X\|P_X)-\lambda H_Q(X)-\nonumber\\
	& &\sum_xQ_X(x)\sum_uQ_{U|X}(u|x)\log[W^\lambda(u)2^{\zeta d(x,u)}
Z^{1+\lambda}(u,x,V,\lambda)]\bigg\}\\
	&=&\inf_{Q_X}\sup_{0\le\lambda\le 1}\inf_{\zeta\ge 0}\min_{V,W}
\bigg\{\zeta D+D(Q_X\|P_X)-\lambda H_Q(X)-\nonumber\\
	& &\sum_xQ_X(x)\min_u\log[W^\lambda(u)2^{\zeta d(x,u)}
Z^{1+\lambda}(u,x,V,\lambda)]\bigg\}\\
	&\dfn&\inf_{Q_X}\sup_{0\le\lambda\le 1}\inf_{\zeta\ge 0}\min_{V,W}
\bigg\{\zeta D+D(Q_X\|P_X)-\lambda H_Q(X)-\nonumber\\
& &\sum_xQ_X(x)\log T(x,\zeta,\lambda,V,W)\bigg\}\\
	&\ge&\sup_{0\le\lambda\le 1}\inf_{\zeta\ge 0}\min_{V,W}\inf_{Q_X}
\bigg\{\zeta D+D(Q_X\|P_X)-\lambda H_Q(X)-\nonumber\\
& &\sum_xQ_X(x)\log T(x,\zeta,\lambda,V,W)\bigg\}\\
	&=&\sup_{0\le\lambda\le 1}\inf_{\zeta\ge 0}\min_{V,W}\inf_{Q_X}\bigg\{\zeta D+
	\sum_xQ_X(x)\log\frac{Q_X^{1+\lambda}(x)}{P_X(x)T(x,\zeta,\lambda,V,W)}\bigg\}\\
	&=&\sup_{0\le\lambda\le 1}\inf_{\zeta\ge 0}\min_{V,W}\inf_{Q_X}\bigg\{\zeta D+\nonumber\\
& &(1+\lambda)\cdot
	\sum_xQ_X(x)\log\frac{Q_X(x)}{[P_X(x)T(x,\zeta,\lambda,V,W)]^{1/(1+\lambda)}}\bigg\}\\
	&=&\sup_{0\le\lambda\le 1}\inf_{\zeta\ge 0}\min_{V,W}\bigg[\zeta D-(1+\lambda)\log\left\{
\sum_x[P_X(x)T(x,\zeta,\lambda,V,W)]^{1/(1+\lambda)}\right\}\bigg]\nonumber\\
&=&\sup_{0\le\lambda\le 1}\inf_{\zeta\ge 0}\min_{V,W}\bigg[\zeta D-(1+\lambda)\log\bigg\{
	\sum_xP_X(x)^{1/(1+\lambda)}\times\nonumber\\
	& &\min_uW^{\lambda/(1+\lambda)}(u)2^{\zeta d(x,u)}\sum_y
        P(y|x)^{1/(1+\lambda)}V(y|u)^{\lambda/(1+\lambda}\bigg\}\bigg]\nonumber\\
&=&\sup_{0\le\lambda\le 1}\inf_{\zeta\ge 0}\min_{C}\bigg[\zeta D-\nonumber\\
	& &(1+\lambda)\log\left\{
	\sum_x\min_u 2^{\zeta d(x,u)}\sum_y
        P(x,y)^{1/(1+\lambda)}C(u,y)^{\lambda/(1+\lambda}\right\}\bigg].
\end{eqnarray}
Concerning eq.\ (\ref{eh}), we have the following chain of equalities and inequalities:
\begin{eqnarray}
	\hat{\bE}(\tR,E_{\mbox{\tiny ecl}},D)&=&\inf_{\{Q_X:~D(Q_X\|P_X)\le E_{\mbox{\tiny ecl}}\}}\sup_M\sup_{\{Q_{U|X}:~-\E _Q\log
	M(U)\le \tR,~{\E}_Qd(X,U)\le D\}}\nonumber\\
	& &\inf_V\sup_{0\le\lambda\le 1}
\bigg\{D(Q_X\|P_X)-\lambda H_Q(X)+\nonumber\\
& &\lambda H_Q(U)-
(1+\lambda)\sum_{u,x}Q_{UX}(u,x)\log Z(u,x,V,\lambda)\bigg\}\nonumber\\
&=&\inf_{\{Q_X:~D(Q_X\|P_X)\le E_{\mbox{\tiny ecl}}\}}\bigg[D(Q_X\|P_X)+\sup_M\sup_{Q_{U|X}}
	\inf_{\rho\ge 0}\inf_{\zeta\ge 0}\inf_V\sup_{0\le\lambda\le 1}
\bigg\{\rho \tR+\zeta D+\nonumber\\
	& &\rho {\E}_Q\log M(U)-\zeta{\E}_Qd(X,U)-\lambda H_Q(X)+\lambda
H_Q(U)-\nonumber\\
& &(1+\lambda)\sum_{u,x}Q_{UX}(u,x)\log
Z(u,x,V,\lambda)\bigg\}\bigg]\nonumber\\
&\ge&\inf_{\{Q_X:~D(Q_X\|P_X)\le E_{\mbox{\tiny ecl}}\}}\bigg[D(Q_X\|P_X)+\sup_M\sup_{Q_{U|X}}
	\sup_{0\le\lambda\le 1}\inf_{\rho\ge 0}\inf_{\zeta\ge 0}\inf_V
\bigg\{\rho\tR+\zeta D+\nonumber\\
	& &\rho {\E}_Q\log M(U)-\zeta{\E}_Qd(X,U)-\lambda H_Q(X)+\lambda
H_Q(U)-\nonumber\\
& &(1+\lambda)\sum_{u,x}Q_{UX}(u,x)\log
Z(u,x,V,\lambda)\bigg\}\bigg]\nonumber\\
&=&\inf_{\{Q_X:~D(Q_X\|P_X)\le E_{\mbox{\tiny ecl}}\}}\bigg[D(Q_X\|P_X)+\sup_{0\le\lambda\le 1}\sup_M\sup_{Q_{U|X}}
	\inf_{\rho\ge 0}\inf_{\zeta\ge 0}\inf_V
\bigg\{\rho \tR+\zeta D+\nonumber\\
	& &\rho{\E}_Q\log M(U)-\zeta{\E}_Qd(X,U)-\lambda H_Q(X)+\lambda
H_Q(U)-\nonumber\\
& &(1+\lambda)\sum_{u,x}Q_{UX}(u,x)\log
Z(u,x,V,\lambda)\bigg\}\bigg]\nonumber\\
&=&\inf_{\{Q_X:~D(Q_X\|P_X)\le E_{\mbox{\tiny ecl}}\}}\bigg[D(Q_X\|P_X)+\sup_{0\le\lambda\le 1}
	\inf_{\rho\ge 0}\inf_{\zeta\ge 0}\inf_V\sup_M\sup_{Q_{U|X}}
\bigg\{\rho\tR+\zeta D+\nonumber\\
	& &\rho{\E}_Q\log M(U)-\zeta{\E}_Qd(X,U)-\lambda H_Q(X)+\lambda
H_Q(U)-\nonumber\\
& &(1+\lambda)\sum_{u,x}Q_{UX}(u,x)\log
Z(u,x,V,\lambda)\bigg\}\bigg]\nonumber\\
&=&\inf_{\{Q_X:~D(Q_X\|P_X)\le E_{\mbox{\tiny ecl}}\}}\bigg[D(Q_X\|P_X)+\sup_{0\le\lambda\le 1}
	\inf_{\rho\ge 0}\inf_{\zeta\ge 0}\inf_V\sup_M\sup_{Q_{U|X}}\inf_W
\bigg\{\rho\tR+\zeta D+\nonumber\\
	& &\rho{\E}_Q\log M(U)-\zeta{\E}_Qd(X,U)-\lambda H_Q(X)-\lambda
\E _Q\log W(U)-\nonumber\\
& &(1+\lambda)\sum_{u,x}Q_{UX}(u,x)\log
Z(u,x,V,\lambda)\bigg\}\bigg]\nonumber\\
&=&\inf_{\{Q_X:~D(Q_X\|P_X)\le E_{\mbox{\tiny ecl}}\}}\bigg[D(Q_X\|P_X)+\sup_{0\le\lambda\le
1}\bigg(-\lambda H_Q(X)+
	\inf_{\rho\ge 0}\inf_{\zeta\ge 0}\bigg[\rho
\tR+\zeta D+\nonumber\\
	& &\inf_V\sup_M\inf_W\sup_{Q_{U|X}}\bigg\{\nonumber\\
	& &\lambda{\E}_Q\log W(U)-\zeta{\E}_Qd(X,U)-
(1+\lambda)\sum_{u,x}Q_{UX}(u,x)\log
Z(u,x,V,\lambda)\bigg\}\bigg]\bigg)\bigg]\nonumber\\
&=&\inf_{\{Q_X:~D(Q_X\|P_X)\le E_{\mbox{\tiny ecl}}\}}\bigg[D(Q_X\|P_X)+\sup_{0\le\lambda\le
1}\bigg(-\lambda H_Q(X)+\nonumber\\
	& &\inf_{\rho\ge 0}\inf_{\zeta\ge 0}\bigg[\rho
\tR+\zeta D+\inf_V\sup_M\inf_W\sup_{Q_{U|X}}\bigg\{\nonumber\\
	& &\sum_xQ_X(x)\sum_uQ_{U|X}(u|x)\log\frac{M^\rho(u)}{W^\lambda(u)2^{\zeta d(x,u)}Z^{1+\lambda}(u,x,V,\lambda)}
\bigg\}\bigg]\bigg)\bigg]\nonumber\\
&=&\inf_{\{Q_X:~D(Q_X\|P_X)\le E_{\mbox{\tiny ecl}}\}}\bigg[D(Q_X\|P_X)+\sup_{0\le\lambda\le
1}\bigg(-\lambda H_Q(X)+\nonumber\\
	& &\inf_{\rho\ge 0}\inf_{\zeta\ge 0}\bigg[\rho
\tR+\zeta D+\nonumber\\
& &\inf_V\sup_M\inf_W\bigg\{
	\sum_xQ_X(x)\max_u\log\frac{M^\rho(u)}{W^\lambda(u)2^{\zeta d(x,u)}Z^{1+\lambda}(u,x,V,\lambda)}
\bigg\}\bigg]\bigg)\bigg]\nonumber\\
&=&\inf_{Q_X}\sup_{\theta\ge 0}
\bigg[D(Q_X\|P_X)+\theta[D(Q_X\|P_X)-E_{\mbox{\tiny ecl}}]+\sup_{0\le\lambda\le
1}\bigg(-\lambda H_Q(X)+\nonumber\\
& &\inf_{\rho\ge 0}\inf_{\zeta\ge 0}\bigg[\rho
\tR+\zeta D+\nonumber\\
& &\inf_V\sup_M\inf_W\bigg\{
	\sum_xQ_X(x)\log\max_u\frac{M^\rho(u)}{W^\lambda(u)2^{\zeta d(x,u)}Z^{1+\lambda}(u,x,V,\lambda)}
\bigg\}\bigg]\bigg)\bigg]\nonumber\\
&\dfn&\inf_{Q_X}\sup_{\theta\ge 0}
\bigg[D(Q_X\|P_X)+\theta[D(Q_X\|P_X)-E_{\mbox{\tiny ecl}}]+\nonumber\\
& &\sup_{0\le\lambda\le
1}\bigg(-\lambda H_Q(X)+
\inf_{\rho\ge 0}\inf_{\zeta\ge 0}
\bigg[\rho
\tR+\zeta D+\nonumber\\
& &\inf_V\sup_M\inf_W
\sum_xQ_X(x)\log S(x,M,W,V,\rho,\zeta,\lambda)
\bigg]\bigg)\bigg]\nonumber\\
&\ge&\sup_{\theta\ge 0}\sup_{0\le\lambda\le 1}\inf_{\rho\ge 0}\inf_{\zeta\ge 0}
\inf_V\sup_M\inf_W\inf_{Q_X}\bigg\{\rho \tR+\zeta D-\theta E_{\mbox{\tiny ecl}}+\nonumber\\
& &(1+\theta)D(Q_X\|P_X)-
\lambda H_Q(X)+\nonumber\\
& &\sum_xQ_X(x)\log S(x,M,W,V,\rho,\zeta,\lambda)
\bigg\}\nonumber\\
&=&\sup_{\theta\ge 0}\sup_{0\le\lambda\le 1}\inf_{\rho\ge 0}\inf_{\zeta\ge 0}
\inf_V\sup_M\inf_W\bigg\{\rho \tR+\zeta D-\theta E_{\mbox{\tiny ecl}}+\nonumber\\
& &(1+\theta+\lambda)]\inf_{Q_X}\sum_{x}Q_X(x)\log\frac{Q_X(x)}{[P_X^{1+\theta}(x)/S(x,M,W,V,\rho,\zeta,\lambda)]^{1/(1+\theta+\lambda)}}
\bigg\}\nonumber\\
&=&\sup_{\theta\ge 0}\sup_{0\le\lambda\le 1}\inf_{\rho\ge 0}\inf_{\zeta\ge 0}
\inf_V\sup_M\inf_W\bigg\{\rho\tR+\zeta D-\theta E_0-\nonumber\\
& &(1+\theta+\lambda)\log
\bigg(\sum_x\bigg[\frac{P_X^{1+\theta}(x)}{S(x,M,W,V,\rho,\zeta,\lambda)}\bigg]^{1/(1+\theta+\lambda)}\bigg)\bigg\}.
\end{eqnarray}

\end{document}